\documentclass{aa}  
\pdfoutput=1
\usepackage{graphicx}
\usepackage{float}
\usepackage{longtable}
\usepackage{txfonts}
\newcommand{\xsurh}[1]{[\rm{#1/H}]}
\newcommand{\xsury}[2]{[\rm{#1/#2}]}
\newcommand{\vsini}{\ensuremath{v_{{\mathrm e}}\sin i}}
\newcommand{\teff}{\ensuremath{T_\mathrm{eff}}}
\newcommand{\logg}{\ensuremath{\mathrm{log}\ g}}
\newcommand{\loggf}{\ensuremath{\mathrm{log}\ gf}}
\newcommand{\kms}{\ensuremath{\mathrm{km.s^{-1}}}}

\newcommand{\xit}{\ensuremath{\xi _{t}}}

\newcommand{\vm}{Varenne \& Monier (1999)}
\begin{document}
\title{Chemical composition of A and F dwarfs members of the Hyades open cluster\thanks{Tables 5 to 8 are only available in electronic format at the CDS.}$^,$\thanks{Based on observations at the Observatoire de Haute-Provence (France).}}

\author{M. Gebran\inst{1}\thanks{Present affiliation: Departament d'Astronomia i Meteorologia, Universitat de Barcelona, c/ Mart\`i i Franqu\`es, 1, 08028 Barcelona, Spain.}, M. Vick\inst{1,2}, R. Monier\inst{3} and L. Fossati\inst{4,5}}

\offprints{M. Gebran}

\institute{$^1$ Groupe de Recherche en Astronomie et Astrophysique du Languedoc,UMR 5024, Universit\'e Montpellier II, Place Eug\`ene Bataillon, 34095 Montpellier, France. \\
\email{mgebran@am.ub.es}\\
$^2$ D\'epartement de Physique, Universit\'e de Montr\'eal, Montr\'eal, PQ, H3C~3J7 \\
\email{mathieu.vick@umontreal.ca}\\
$^3$Laboratoire Universitaire d'Astrophysique de Nice, UMR 6525, Universit\'e de Nice - Sophia Antipolis, Parc Valrose, 06108 Nice Cedex 2, France.\\
\email{Richard.Monier@unice.fr}\\
$^4$Institut f\"ur Astronomie, Universit\"at Wien, T\"urkenschanzstrasse 17, 1180 Wien, Austria.\\
\email{fossati@astro.univie.ac.at}\\
$^5$Department of Physics and Astronomy, Open University, Walton Hall, Milton Keynes MK7 6AA, UK.\\
\email{l.fossati@open.ac.uk}}
\date{Received ; accepted }

 
  \abstract
   {}
   {Abundances of \textbf{15} chemical elements have been derived for 28 F and 16 A stars
   members of the Hyades open cluster in order to set constraints on
   self-consistent evolutionary models 
   including radiative and turbulent diffusion.}
   {A spectral synthesis iterative procedure was applied to derive the
   abundances from selected high quality lines in high resolution high
   signal-to-noise spectra obtained with SOPHIE and AURELIE at the Observatoire
   de Haute Provence. } 
     {The abundance patterns found for A and F stars in the Hyades resemble
     those observed in Coma Berenices and Pleiades clusters. In graphs
     representing the abundances versus the effective temperature, A stars
     often display abundances much more scattered around their mean values than
     the coolest F stars do. Large star-to-star variations are detected in  
     the Hyades A dwarfs in their abundances of
     C, Na, Sc, Fe, Ni, Sr, Y and Zr, which we interpret as evidence of
     transport processes competing with radiative diffusion.\\
    In A and Am stars, the abundances of Cr, Ni,
    Sr, Y and Zr are found to be correlated with that of iron as in the Pleiades and in Coma Berenices. The ratios \xsury{C}{Fe} and
    \xsury{O}{Fe} are found to be anticorrelated with 
        \xsurh{Fe} as in Coma Berenices. All Am stars in the Hyades are
	deficient in C and O and overabundant in elements heavier than Fe but
	not all are deficient in calcium and/or scandium. The F stars have solar
	abundances for almost all elements except for Si.\\
         The overall shape of the abundance pattern of the slow rotator HD30210 cannot be entirely
	 reproduced by models including radiative diffusion and different amounts of turbulent
	 diffusion.}
   {While part of the discrepancies between derived and predicted abundances
   could be due to non-LTE effects, the inclusion of competing processes such
   as rotational mixing and/or mass loss seems necessary in order to improve the agreement between the observed and predicted abundance patterns.}


\authorrunning{Gebran et al.}
\maketitle

\section{Introduction}
Abundance determinations for A and F dwarfs in open clusters and moving groups
aim at elucidating the mechanisms of mixing at play in the interiors of these
main-sequence stars.
This paper is the third in a series addressing the chemical composition of A and F dwarfs in open clusters of different ages. 
The objectives of this long-term project are twofold: first, 
we wish to improve our knowledge of the chemical composition of A and F dwarfs, and secondly we aim to use these determinations to set constraints 
on particle transport processes in self consistent evolutionary models.
The first paper (Gebran et al. 2008, hereafter Paper I) addressed the abundances of several chemical elements for 11 A and 11 F 
dwarf members of the Coma Berenices open cluster. In the second paper (Gebran \& Monier 2008, Paper II),  abundances were derived  for the same chemical elements 
for 16 A and 5 F dwarf members of the Pleiades open cluster. In this study, we present a re-analysis of the A and F dwarf abundances in the Hyades open cluster, already addressed
by Varenne \& Monier (1999) using mono-order 
spectra on much more limited spectral ranges (three 70 \AA\ wide spectral
intervals). The new data we collected are high signal to noise and high resolution 
\'echelle spectra stretching over more than 3000 \AA\ which enabled us to synthesize more lines with high quality atomic data (and more elements) than in Varenne \& Monier (1999).
In this paper, abundances have been derived for \textbf{15 chemical elements (C, O, Na, Mg, Si, Ca, Sc, Ti, Cr, Mn, Fe, Ni, Sr, Y and Zr)} for 28 F and 16 A  members of the Hyades cluster.\\

Open clusters are excellent laboratories to test stellar evolution theory. Indeed stars in open clusters originate from the same interstellar material, 
and thus have the same initial chemical composition and age. 
At a distance of $\sim$46 pc (van Leeuwen 2007), the Hyades open cluster is the
nearest star cluster and also the most analyzed of 
all clusters. Perryman et al. (1998) compared the observational HR diagram of the Hyades with stellar evolution models and obtained an estimation of the age 
of this cluster ($\sim$625 Myr) using a combination of Hipparcos data with ground-based photometric indexes. 
Boesgaard \& Friel (1990) derived a metallicity for the Hyades slightly above
solar ( $<$\xsurh{Fe}$>$=0.127$\pm$0.022 dex ) from their analysis of Fe I lines
in 14 F dwarfs. In a study of 40 Hyades G dwarfs, \cite{1997ESASP.402..687C}
also derived a mean 
 metallicity of $<$\xsurh{Fe}$>$=+0.14$\pm$0.05 dex. Abundances derived from
 calibration of Geneva photometry by \cite{2000IAUJD..13E...7G} ($<$\xsurh{Fe}$>$=+0.14$\pm$0.01) 
also yield a slightly enhanced metallicity.\\

Several papers have addressed the chemical composition of A and F dwarfs in the Hyades open cluster. Carbon and iron abundances have been derived for 14 F stars 
by \cite{1990ApJ...351..480F} and \cite{1990ApJ...351..467B}. Lithium abundances have been determined for several F, G and K dwarfs by 
Cayrel et al. (1984), \cite{1986ApJ...303..724B}, \cite{1988ApJ...332..410B} and \cite{1993ApJ...415..150T}. \cite{1993ApJ...412..173G} have derived the oxygen 
abundances for 26 F dwarfs members of the Hyades cluster. Carbon, oxygen, sodium, magnesium, silicon, calcium, scandium, chromium, iron, nickel, yttrium and 
barium abundances have been derived for A stars by \cite{1997PASJ...49..367T}, \cite{hui-et-alecian98}, \cite{burkhart-coupry-2000} and \vm. Most of these studies, 
at the exception of \vm, have focused mainly on the peculiar Am stars, leaving
aside the normal A stars. For a given chemical element, 
large star-to-star variations were found among A stars in several open clusters like the Pleiades (Paper II) and Coma Berenices (Paper I) 
. \vm \ found significant star-to-star variations in the abundances of O, Na, Ni, Y and Ba for A stars in the Hyades whereas the F dwarfs display much less dispersion.
 Similarly, star-to-star variations of [Fe/H], [Ni/H] and [Si/H] are larger for the A dwarfs than for the F dwarfs in the Ursa Major group (Monier 2005). 
 This behavior was also observed in earlier works on field A stars (Holweger et al. 1986, Lambert et al. 1986, Lemke 1998, 1990, Hill \& Landstreet 1993, Hill 1995, Rentzsch-Holm 1997 
 and Varenne 1999). \\
 The incentive to reanalyze the chemical composition of A and F dwarfs of the Hyades is justified by the acquisition of higher quality spectra encompassing a much wider spectral
 range than used in Varenne \& Monier (1999). This allowed us to model more
 lines of higher quality (ie. with more accurate atomic data) for most investigated chemical elements yielding more accurate abundances for several species.

 We have also searched for correlations of the abundances of individual elements with that of iron, an issue not addressed in Varenne \& Monier (1999). Furthermore, the state of the art of modelling the internal structure and evolution of A dwarfs has improved over the last ten years and we present here comparisons of new models to the observed pattern of abundances.\\
The selection of the stars and the data reduction are described in \S \ref{sec:obs-reduc}. The determination of the fundamental parameters (\teff \ and \logg) and the spectrum synthesis computations are discussed in (\S \ref{sec:spectra}). As in Papers I and II, the behavior of the abundances of the analysed chemical elements  in A and F dwarfs have been investigated in \S
\ref{sec:results} with respect to effective temperature (\teff), projected rotational velocity (\vsini) and the iron abundance (\xsurh{Fe}).  
In \S \ref{sec:models}, the found abundance patterns are compared to recent evolutionary models including self consistent 
treatment of particle transport (Turcotte et al. 1998b; Richer et al. 2000 and Richard et al. 2001). The abundance pattern of the Am star, HD 30210,
is modelled in detail using the latest prescriptions in the Montreal code (Richer et al., 2000).
Conclusions are gathered in \S \ref{sec:conclusion}. 
%



\section{Program stars, observations and data reduction}
\label{sec:obs-reduc}
\begin{figure}[h]
\centering
\includegraphics[scale=0.35]{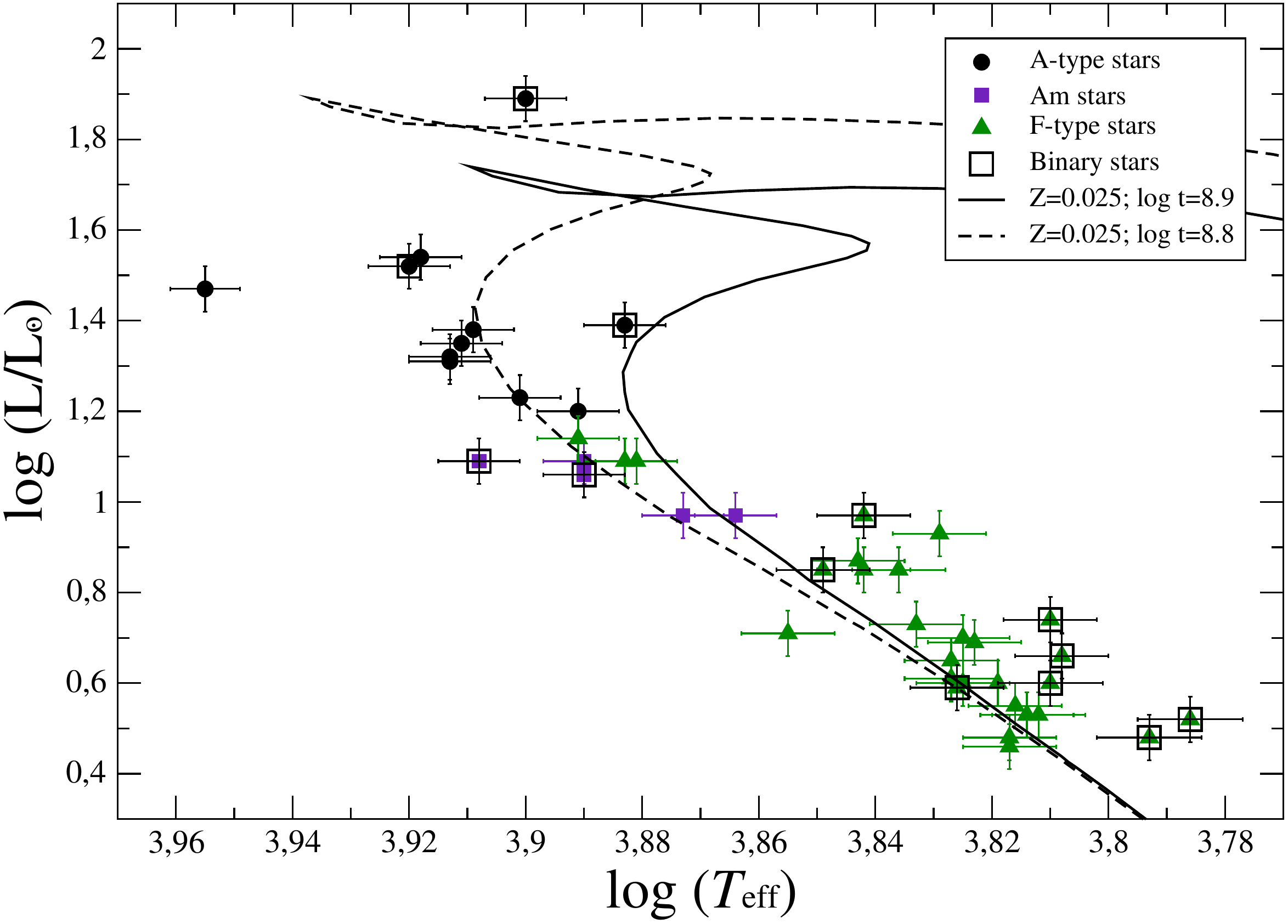}
\caption{HR diagram of the Hyades cluster. The two isochrones are calculated with the age given in
WEBDA (\ensuremath{\log t}=8.9 - full line) and our best fit (\ensuremath{\log t}=8.8 - dashed line.) }
\label{fig:hr}
 \end{figure}

Our observing sample consists of 28 F and 16 A members of the Hyades cluster
brighter than V=7 (the same
sample selected by Varenne \& Monier 1999). At the distance of the Hyades, V=7 mag 
corresponds to the latest F dwarfs (F8-F9). 
A Hertzsprung--Russell (HR) diagram of the Hyades, shown in Fig.\ref{fig:hr}, was constructed using the effective temperatures we derived in section 3
, the $V$ magnitudes retrieved from SIMBAD and appropriate bolometric corrections.
We adopted a cluster distance of 46.5$\pm$0.3 pc (van Leeuwen 2007), a reddening of 0.010$\pm$0.010 mag (WEBDA\footnote{www.univie.ac.at/webda/}) and the bolometric 
corrections given by \cite{balona}. The uncertainty on the bolometric correction is of the order of 0.07 mag, leading to a typical uncertainty in $M_{\mathrm{bol}}$ of 
about 0.15 mag, corresponding to an uncertainty of about 0.05 dex in \ensuremath{\log L/L_{\odot}}. 
In the HR diagram of Fig. \ref{fig:hr} the Am stars are depicted as filled squares,
the normal A stars as filled circles and the F stars as filled triangles, the spectroscopic binaries as open squares. We did not 
 correct the luminosities of the binaries since the contribution to the total flux due to the secondary is not known. 
 
The age of the cluster can be estimated by adjusting isochrones by \cite{marigo08}, including overshooting and calculated for a metallicity of 
Z = 0.025 dex, which corresponds roughly to the mean of the recent determinations by \cite{perryman98,castellani02,percival03,salaris04,taylor06,holmberg07} 
plus the value given in WEBDA. Fig.\ref{fig:hr} displays two isochrones: one corresponding to the age given in WEBDA (\ensuremath{\log t} = 8.9 - full line) and our 
best fit corresponding to a slightly lower age (\ensuremath{\log t} = 8.8 - dashed line), which agrees quite nicely with the determination by 
\cite{perryman98}: t=625$\pm$50 Myr. We are inclined to rule out a lower metallicity which would lead to a much lower value of the age (models without overshooting 
would lead to a much lower age). Following \cite{land2007}, we have also derived M/M$_{\odot}$, fractional age (fraction of time spent on the Main Sequence noted as $\tau$) and their uncertainties for each star and collected them in online Table
\ref{tab:LTMA}. In Fig. \ref{fig:hr}, the star HD 27962 (= 68 Tau) is located at a much higher
effective temperature than the other A dwarfs of similar luminosities.
Mermilliod (1982) confirmed its membership to the Hyades and proposed that HD
27962 is a blue straggler with the spectral characteristics of an Am star. Abt
(1985) assigned a spectral type Am (A2KA3HA5M) to HD 27962. 

\vskip0.7 cm

The A stars were observed using 
SOPHIE, the \'echelle spectrograph at the Observatoire de Haute-provence (OHP). SOPHIE spectra stretch from 3820 to 6930 \AA \ in 39 
orders with two different spectral resolutions: the high resolution mode HR (R=75000) and the high efficiency mode HE (R=39000). All A stars were observed in the HR mode.
The observing dates, exposure times and Signal to Noise ratios achieved for each
A star are collected in Table \ref{tab:obs}.
On good nights, an exposure time of 25 minutes typically yielded well exposed spectra with signal-to-noise ratios ranging
from 300 to 600, depending on the V magnitude. 
As we did not get enough observing time to observe the F stars with SOPHIE as
well, we have used the mono-order AURELIE spectra obtained by \vm. For each F
star, three spectral regions centred on $\lambda$6160 \AA, $\lambda$5080 \AA \ and 
$\lambda$5530 \AA \ had been observed at resolutions 30000  (V $\geq$ 6) or
60000 (V $\leq$ 6) and signal to noise ratios close to 200.\\
The fundamental data for the selected stars are collected in Table \ref{A-F-stars}. The van Bueren and Henry Draper identifications appear in columns 1 and 2, 
the spectral type retrieved from SIMBAD or from Abt \& Morell (1995) in column 3 and the apparent magnitudes
in column 4. Effective temperatures (\teff) and surface gravities (\logg), 
derived from $uvby\beta$ photometry (see section 3), appear in columns 5 and 6. 
The derived projected rotational velocities and the microturbulent velocities
are in columns 7 and 8.
Comments about binarity and pulsation appear in the last column. The apparent rotational velocities range 
from 11 \kms \ to 165 \kms, only 7 stars rotate faster than 100 \kms.\\
Inspection of the CCDM catalogue (Dommanget \& Nys 1995) reveals that 13 among the 16 A stars are in binary or multiple systems. All of these stars are primaries and the components are much fainter (1 mag$<\Delta m<$9.5 mag ). Only the case of \textbf{HD27962} (CCDM J04255+1755 AB) has to be accounted for because its companion (component B) is three magnitudes fainter than A and at only 1.4" from A.
In case of F stars, 13 among the 28 stars belong to multiple systems. Six of them have nearby companions whose angular distance was probably less than the fiber angular size on the sky (3 arcsec) and who are only three magnitudes fainter than the brightest star we analysed. The spectral types of these companions is unknown. For the F stars, these are:\\
- \textbf{HD26015} = CCDM J04077+1510A  has a companion at about 4" with $\Delta m$ = 2.8 mag \\
- \textbf{HD27383} = CCDM J04199+1631AB: components A and B are very close with $\Delta m = 2.0$ mag \\
- \textbf{HD27991} = CCDM J04257+1557AP: companion P is quoted to be at 0.1" with $\Delta m = 0.7$ mag\\
- \textbf{HD28363} = CCDM J04290+1610AB is a spectroscopic binary whose angular separation is not specified with a $\Delta m = 1.0$ mag \\
- \textbf{HD30810} = CCDM J04512+1104AB is a triple star whose component B has same magnitude as A, no angular separation is provided. \\
For these six stars, we believe that the light of the companions might have contaminated the spectra of the brightest components we analysed. The effects are probably most pronounced for the F stars HD26015, HD27383, HD27991, HD28363 and HD30810. \\
The SOPHIE spectra were reduced using IRAF (Image Reduction and Analysis Facility, Tody 1993) in order to properly correct for scattered light. The sequence of IRAF procedures, which follows the method devised by \cite{2002A&A...383..227E}, is fully described in Paper I.

\begin{table*}
\caption{Basic physical quantities for the programme stars.}
\label{A-F-stars}
\begin{center}
\begin{tabular}{c c c c c c c c c}
\hline
   vB          &  HD & Type & $m_{v}$ & \teff & \logg & \vsini & \xit & Remarks \\
               &     &      &         & (K)   &  ($\mathrm {cm.s^{-2}}$) &  (\kms)     & (\kms) &\\
\hline
\multicolumn{9}{c}{A stars} \\ \hline
38   & 	27628 	& A3m   & 5.72&7310 & 4.12 & 31.2 & 3.70 & $\delta$ Scuti (d) \\
47   &	27819 	& A7V/A8V 	&4.80 &8190 & 3.94 & 47.0 & 3.00 & \\
54   & 	27934 	& A7IV/A6V 	&4.22 &8290 & 3.83 & 80.0 & 3.00 & \\
56   & 	27962 	& A2IV/Am 	& 4.29&9025 & 3.95 & 11.3 & 2.80 & Blue Straggler\\
67   & 	28226 	& A3m	&5.72 &7465 &4.09 & 83.0 & 3.30 & \\
72   & 	28319	& A7III/A7IV	& 3.39&7950 & 3.70 & 65.0 & 2.70 & SB (b), $\delta$Scuti (d) \\
74   & 	28355	& A7V/A5m 	& 5.03&7965 & 3.97 & 90.0 & 3.00 & \\
82   & 	28527 	& A6IV/A7V 	&4.78 &8180 & 3.98 & 67.8 & 3.70 &\\
83   & 	28546 	& A5m 	& 5.48&7765 & 4.20 & 27.5 & 3.80 & SB (b)\\
95   & 	28910 	& A8V/A7V 	& 4.65&7640 & 4.02 & 110.0 & 2.60 & SB (b), $\delta$Scuti (d)\\
104  &  29388 	& A6V 	&4.27 &8310 & 3.87 & 81.0 & 3.30 & SB (b)\\
107  & 	29499 	& A5m/A9III 	& 5.39&7770 & 4.11 & 60.0 & 3.00 & \\
108  &  29488 	& A5V/A6V 	&4.70 &8150 & 3.80 & 118.0 & 2.60 & \\
112  & 	30210 	& A2m 	& 5.37&8080 & 3.92 & 57.0 & 4.00& SB (b)\\
123  & 	30780 	& A7V/A9V 	&5.11 &7790 & 3.90 & 165.0 & 2.80 & $\delta$ Scuti (d)\\
129  &  32301 	& A7V 	& 4.64&8110 & 3.73 & 117.0 & 2.70 & \\
   &&&&&&&& \\ \hline
   \multicolumn{9}{c}{F stars} \\ \hline
6   & 24357 & F4V	 &5.97 &6975 & 4.13 & 65.5 & 2.10 &  \\
8   & 25102 & F5V	 & 6.37&6685 & 4.32 & 64.0 & 2.20 &\\
11  & 26015 & F3V	 &6.01 &6860 & 4.46 & 29.7 & 1.80 &  \\
13  & 26345 & F6V 	&6.62 &6685 & 4.36 & 27.2 & 1.60 &   \\
14  & 26462 & F4V 	& 5.73&6945 & 4.14 & 18.6 & 1.50 & SB (b)\\
20  & 26911 & F5V 	&6.32 &6810 & 4.26 & 62.1 & 2.20 &\\
29  & 27383 & F9V	 &6.88 &6215 & 4.39 & 16.8 & 1.40 & SB (b)\\
33  & 27459 & F0V/F0IV 	&5.26&7785 & 3.98 & 73.0 & 3.10 &$\delta$ Scuti (d)\\
35  & 27524 & F5V 	&6.80 &6515 & 4.24 & 72.5 & 2.10 & \\
36  & 27534 & F5V	 & 6.80&6485 & 4.26 & 48.0 & 1.50 & \\
37  & 27561 & F5V 	& 6.61&6710 & 4.35 & 22.2 & 1.60 & \\
51  & 27848 & F8 	 &6.97 &6565 & 4.33 & 37.7 & 1.50 &  \\
57  & 27991 & F7V	 &6.46 &6430 & 4.48 & 17.6 & 1.10 & SB2 (c), SB(b)\\
75  & 28363 & F8V 	& 6.59&6325 & 4.43 & 16.0 & 1.00 & SB3 (c), SB (b)\\
78  & 28406 & F6V	 &6.92 &6560 & 4.40 & 33.0 & 1.60 &  \\
84  & 28556 & F0V	 & 5.41&7635 & 4.07 & 83.5 & 3.30 &\\
85  & 28568 & F2 	&6.51 &6710 & 4.40 & 64.0 & 1.90 & \\
89  & 28677 & F4V/F2V 	&6.02 &7060 & 4.07 & 129.0 & 1.90 & SB (b)\\
90  & 28736 & F5V 	& 6.40&6655 & 4.30 & 48.0 & 1.70 & \\
94  & 28911 & F2 	&6.62 &6590 & 4.32 & 50.0 & 1.60 & \\
100 & 29169 & F5IV 	&6.02 &6950 & 4.25 & 75.0 & 2.20 & \\
101 & 29225 & F8 	&6.65 &6700 & 4.41 & 50.8 & 1.70 & SB (b)\\
111 & 30034 & F0V/A9IV 	&5.40 &7600 & 4.05 & 100.0 & 2.60 &\\
122 & 30810 & F6V 	&6.76 &6110 & 4.36 & 12.0 & 1.10 & SB2 (c), SB (b)\\
124 & 30869 & F5  	& 6.25&6460 & 4.28 & 23.5 &1.80 & SB2 (a,c), SB (b) \\
126 & 31236 & F3IV/F1V 	& 6.37&7165 & 3.89 & 120.0 & 3.00 &\\
128 & 31845 & F5V 	& 6.75&6550 & 4.37 & 33.7 & 1.90& \\
154 & 18404 & F5IV 	& 5.80&6740 & 4.37 & 26.5 & 1.70 & \\  
 & Procyon&F5IV-V&  0.34&6650&4.05&6.0&2.2 & \\
\hline
\multicolumn{9}{l}{\scriptsize
References (a) (b) (c) and (d) are for Griffin et al. (1985), Perryman et al. (1998), Barrado \& Stauffer (1996) and Solano \& Fernley }\\
\multicolumn{9}{l}{\scriptsize
(1997) respectively.}
\end{tabular}
\end{center}
\end{table*}

\begin{table}[h]
\caption{Observing log for the A stars of the Hyades open cluster.}
\label{tab:obs}
\centering
\begin{tabular}{ccccc}
\hline  \hline
HD&spectral &exposure & S/N &Date\\ 
& type  & time (s)&& \\ \hline
 27628   & A3m       & 1500	 &350	  &   10/04/06  	      \\ 
 27819   & A7V       &  1200	&498	  &   10/04/06        \\  
 27934   & A7V       &900	 &591	  &   10/04/06         \\	  
 27962   & A2IV      &  600    &429	 &    10/04/06        \\ 
 28226   & A3m       &1500	&380	  &   10/04/06        \\ 
 28319   & A7III     &  300    &429	 &    10/04/06        \\ 
 28355   & A7V       &  1500	&427	  &   10/04/06        \\ 
 28527   & A6IV      &  1200	&354	  &   10/04/06        \\ 
 28546   & A5m       &  1500	&292	  &   10/04/06        \\ 
 28910   & A8V       &  1200	&434	  &   10/04/06        \\ 
 29388   & A6V       &  600    &561	 &    10/05/06        \\ 
 29499   & A5m       &  1320	&476	&     10/05/06  	      \\ 
 29488   & A5V       &  900    &601	 &    10/05/06        \\ 
 30210   & Am	     &  1320	&474	  &   10/05/06        \\ 
 30780   & A7V       &  1200	&549	  &   10/05/06        \\ 
 32301   & A7V       &  720    &526	 &    10/05/06        \\ 
 \hline		
 \hline	
\end{tabular}
\end{table}

\section{Abundance analysis: method and input data}
\label{sec:spectra}
The abundances of \textbf{15} chemical elements have been derived by iteratively adjusting synthetic spectra to the normalized spectra and minimizing the chi-square of the models to the observations.
Spectrum synthesis is mandatory as the apparent rotational velocities range from 11 to 165 \kms. Specifically, synthetic spectra were computed assuming LTE using Takeda's (1995) iterative procedure and double-checked using Hubeny \& Lanz (1992) SYNSPEC48 code. This version of SYNSPEC calculates lines for elements up to Z=99.
\subsection{Atmospheric parameters and model atmospheres}
The effective temperatures and surface gravities were determined using the UVBYBETA code developed by \cite{1993A&A...268..653N}. This code is based on the \cite{1985MNRAS.217..305M}'s grid, which calibrates the $uvby\beta$ photometry in terms of \teff \ and \logg. The photometric data were taken from \cite{1998A&AS..129..431H}. The estimated errors on \teff \ and \logg, are $\pm$125 K and $\pm$0.20 dex, respectively (see Sec. 4.2 in Napiwotzki et al. 1993).
The found effective temperatures and surface gravities are collected in table 1.\\
The ATLAS9 (Kurucz 1992) code was used to compute LTE model atmospheres assuming
a plane parallel geometry, a gas in hydrostatic and radiative equilibrium and
LTE. The ATLAS9 model atmospheres contain 64 layers with a regular increase in
$\log \tau_{Ross} = 0.125$ and were calculated assuming \cite{1998SSRv...85..161G} solar chemical composition.
This ATLAS9 version uses the new Opacity Distribution Function (ODF) of Castelli \& Kurucz (2003) computed for that solar chemical composition. Convection is calculated in the frame of the mixing length theory (MLT). We have adopted Smalley's prescriptions (\cite{2004IAUS..224..131S}) 
fo the values of the ratios of the mixing length to the pressure scale height ($\alpha=\frac{L}{H_{P}}$) and the microturbulent velocities (constant with depth).

\subsection{The linelist \label{linelist}}
\textbf{For the A stars, the linelist used for spectral synthesis is the same as in Paper I.
All transitions between 3000 and 7000 \AA\ from Kurucz's gfall.dat\footnote{http://kurucz.harvard.edu/LINELISTS/GFALL/} linelist were selected. The abundance analysis relies on more than 200 transitions for the 15 selected elements as explained in Paper I. The adopted atomic data for each elements are collected in Table 8 of Paper I where, for each element, the wavelength, adopted oscillator strength, its accuracy (when available) and original bibliographical reference are given.}
The two sodium lines at 5890 and 5896 \AA \ were not used in this present paper. These lines are likely to be affected by \textbf{interstellar absorption and} non-LTE effects, an LTE treatment assuming depth independent microturbulence underestimates abundances (Takeda et al. 2009).  
For the F stars, the same linelist was used except for iron and magnesium since
only FeI and MgI \textbf{lines} are available in the AURELIE spectra. We have used the
same atomic data for these lines as those in Table 3 of \vm. 
Most of the lines studied here are weak lines formed deep in the atmosphere where
LTE should prevail. They are well suited for abundance determinations. \textbf{We have also included data for hyperfine splitting for the selected transitions when relevant, using the linelist gfhyperall.dat\footnote{http://kurucz.harvard.edu/LINELISTS/GFHYPERALL/}. However the moderate spectral resolution of the spectra and smearing out of spectra by stellar rotation clearly prevent us from detecting signatures of hyperfine splitting and isotopic shifts in our spectra.}

\subsection{Spectrum synthesis}

For each modelled transition, the abundance was derived iteratively using
Takeda's (1995) procedure which minimizes the chi-square between the 
normalized synthetic spectrum and the observed one. \textbf{As explained in Paper I, Takeda's code consists in two routines. The first routine computes the opacity data and it is based on a modified version of Kurucz's Width9 code (\cite{1992RMxAA..23...45K}) while the second computes the normalized flux and minimizes the dispersion between synthetic and observed spectra (see Paper I for a complete
description of the method).\\
We first derived the rotational ($v_{e}\sin i$) and microturbulent ($\xi_{t}$) velocities using several weak and moderately strong FeII lines located between 4491.405 \AA \ and 4508.288 \AA \ and the MgII triplet at 4480 \AA\ by allowing small variations around solar abundances of Mg and Fe as explained in Sec. 3.2.1 of Paper I. The weak iron lines are very sensitive to rotational velocity but not to microturbulent velocity while the moderately strong FeII lines are affected mostly by changes of microturbulent velocity. The MgII triplet is sensitive to both $\xi_{t}$ and $v_{e}\sin i$. Once the rotational and microturbulent velocities were fixed, we then derived the abundance that minimized the chi-square for each transition of a given chemical element. These individual abundances may differ because of different levels of accuracies in the atomic data of each line and possibly because of deviations from LTE in a few of them. The derivation of the mean abundance from these individual abundances is explained in Sec.~\ref{sec:meanuncer}.} The abundances were then double-checked using Hubeny \& Lanz's (1992) SYNSPEC48 code. 

\begin{figure}[h!]
\centering
\includegraphics[scale=0.34]{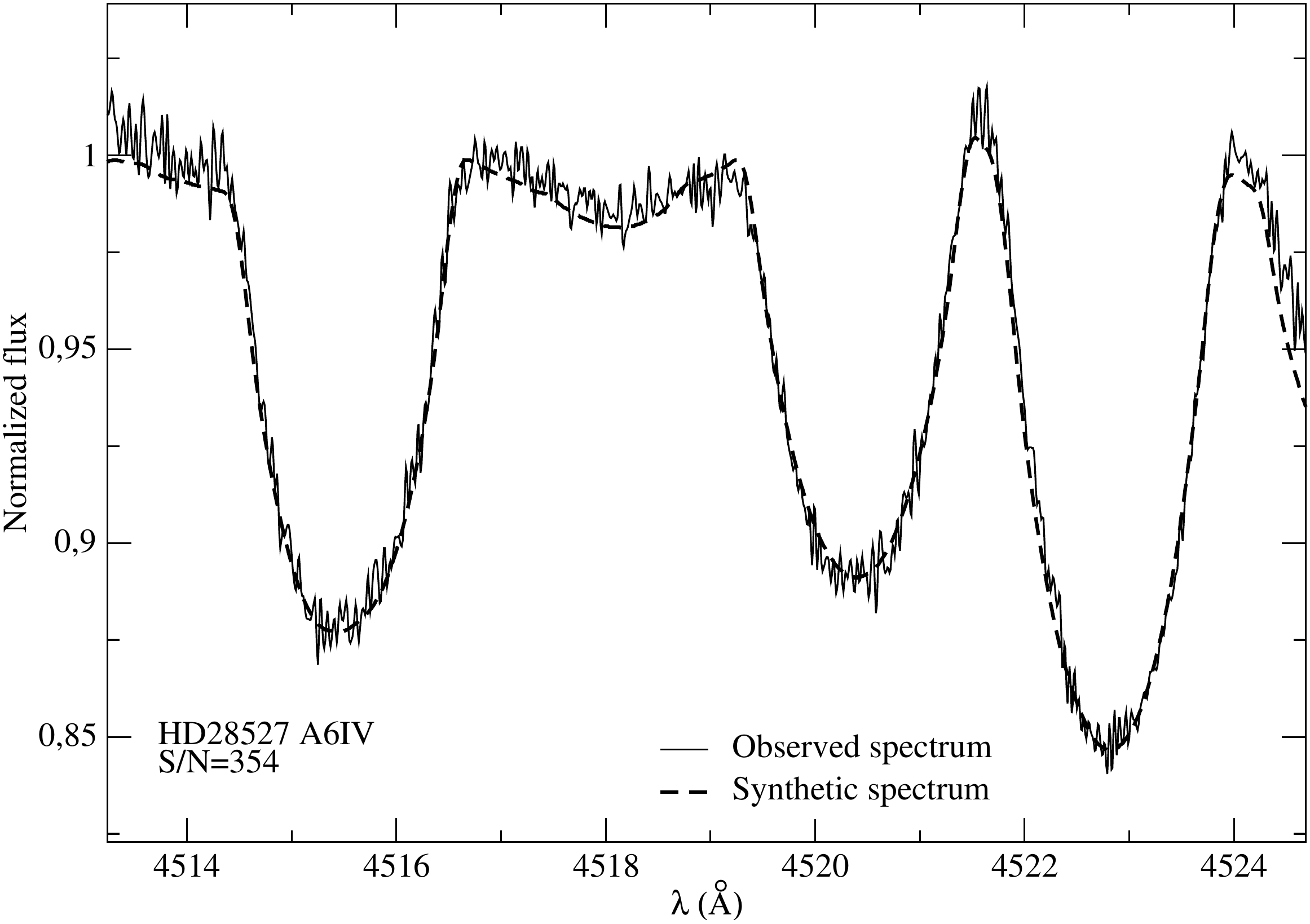}
\caption{A typical agreement between the observed spectrum (thin line) of HD28527 (A6IV) and the synthetic spectrum (dashed thick line) computed as explained in Sect. \ref{sec:spectra}. Three iron (FeII) lines calculated for [Fe/H]=+0.30 dex, the mean iron abundance, are displayed in this figure.}
\label{fig:ajustement}
 \end{figure}

\textbf{As an example, we display in Fig.~\ref{fig:ajustement}  the final synthetic spectrum which best fits several Fe II lines in the observed spectrum of HD28527 (A6IV) in the spectral interval 4513-4525 \AA. In this region, the independent fit of each line yields only slightly different abundances. The displayed synthetic spectrum is computed for an iron abundance of +0.30 dex, which is the derived mean value in HD28527.\\
In the case of stars rotating faster than about 80 km/s, the following neighbouring lines blend: Ti II 4394.059 \AA\ and Ti II 4395.051 \AA, Mn I 4033.062 \AA\ and 4034.483 \AA\ and for the O I lines "triplet" at 6155.900 \AA, 6156.750 \AA\ and 6158.1 \AA.  In these cases, the abundances given in the electronic Table~\ref{element-abundances} are those that provide the best match to each blend.}

\subsection{Internal Consistency checks on the spectral energy distribution}

\begin{figure}[h]
\centering
\includegraphics[scale=0.34]{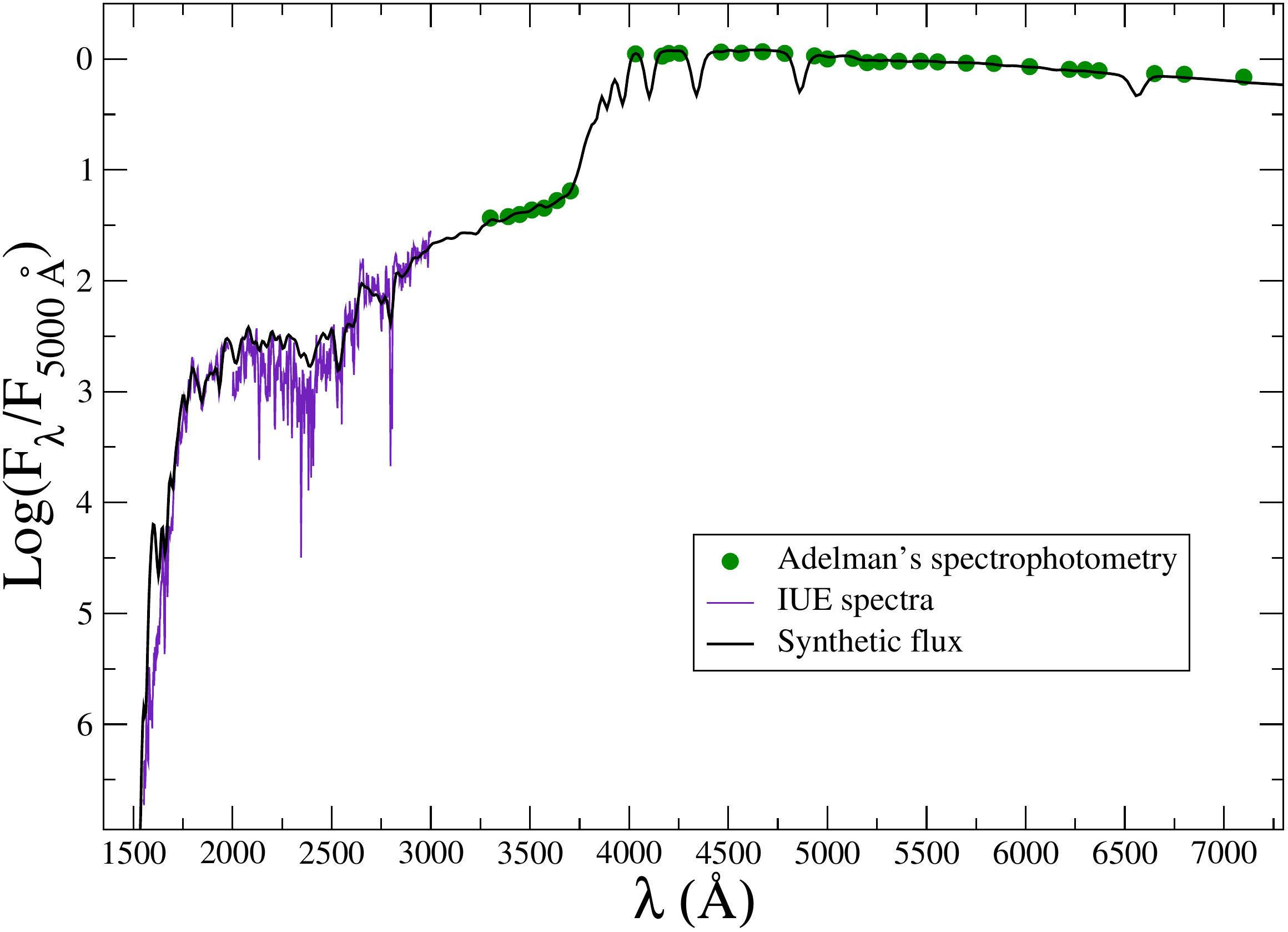}
\caption{Comparison of the LLmodel theoretical spectral energy distribution of HD27819 with the observed spectrophotometry of HD 27819 in the UV and the optical.}
\label{fig:hd27819}
 \end{figure}

For a few stars, we have checked the fundamental parameter determinations by
modelling the entire spectral energy distribution from far UV to IR with the
theoretical ATLAS9 flux computed for the derived fundamental parameters and the
individual abundances. Fig. \ref{fig:hd27819} exemplifies this check for HD
27819. The theoretical spectral energy distribution was computed with the
LLmodel code (Shulyak et al. 2004). The observed spectral energy distribution
was constructed from Adelman et al's (1989) spectrophotometry and IUE
spectrophotometry: SWP04446 (low resolution) + LWP16605 (high resolution
resampled to the LWP low resolution).
The theoretical LLmodel spectral energy distribution, degraded to a spectral
resolution comparable to that of the IUE low resolution spectra, follows nicely the overall
shape of the observed flux distribution which leads credence to the adopted
fundamental parameters and the derived abundances.

\subsection{Mean abundances and uncertainties}
\label{sec:meanuncer}
Apparent rotational velocity, microturbulent velocity were determined for all sample stars. The abundances of \textbf{15} chemical elements were determined for 
most of the stars (when the selected lines were accessible with good signal-to-nois ratios). The abundances 
for A and F stars are collected in Online Tables \ref{abondances-A} and \ref{abondances-F}. 
These abundances are relative to the sun\footnote{($[\frac{X}{H}]=\log(\frac{X}{H})_{\star}-\log(\frac{X}{H})_{\odot}$)}. 
Solar abundances are from Grevesse \& Sauval (1998).
\textbf{For each chemical element the final abundance is an average of the abundances derived from each line. The errors on the final abundances (labelled as $\sigma$) are standard deviation assuming a Gaussian distribution of the abundances derived from each line:} 
\begin{equation}
\bar{x}=\frac{\sum_{i}^{}x_{i}}{N}
\end{equation}
\begin{equation}
\sigma^{2}=\frac{\sum_{i}^{}(x_{i}-\bar{x})^{2}}{N}
\end{equation}
\textbf{where $\bar{x}$ is the mean value of the abundance, N the number of lines of the element and $\sigma$ the standard deviation.\\
Accordingly the error on the abundance of a given element depends on the individual abundances derived from each line and usually varies from star to star. When only one transition was used to derive the abundance (for CI, OI\footnote{Oxygen lines are blended in F stars because of the low resolution of AURELIE spectra, which is not the case of SOPHIE's A stars.}, MgI, SiII, CaII, ScII and YII in F stars), the corresponding error was computed according to the formulation explained in Appendix A of Paper I. It consists into perturbating each of the 6 nominal parameters ($T_{\rm{eff}}$, $\logg$, $\xi_{t}$, $v_{e}\sin i$, $\log gf$ and the continuum position) and repeating the fit for each line. The perturbations $\Delta$(\teff) and $\Delta$(\logg) are 200 K and 0.20 dex, respectively (Napiwotzki et al. 1993). $\Delta$(\vsini) is estimated as 5\% of the nominal $v_{e}\sin i$ and $\Delta$(\xit) is 1 \kms \ (Gebran 2007). $\Delta$(\loggf) depends on the accuracy of the considered lines. It varies from 3\% to more than 50\%. For more details concerning the accuracies on the oscillator strengths, see Tab. 8 (3$^{rd}$ column) of Paper I and Tab. 3 (3$^{rd}$ column) of \vm. The continuum placement error depends on the rotational velocity of the star and is fully explained in Paper I. \\
The difference between the nominal abundance and the one derived with the perturbated parameter yields the uncertainty affected to the given parameter. Considering that the errors are independent, the upper limit of the total uncertainty $\sigma_{tot_{i}}$ for a given transition (i) is:}\\
\begin{equation}
\sigma_{tot_{i}}^{2}=\sigma_{T_{\rm{eff}}}^{2}+\sigma_{\log g}^{2}+\sigma_{\xi_{t}}^{2}+\sigma_{v_{e}\sin i}^{2}+\sigma_{\log gf}^{2}+\sigma_{cont}^{2}.
\end{equation}  
\textbf{Online Table 8 collects the abundances derived for each transition for each studied element in all A and F stars including Procyon (F5V) which served as control star for the spectral synthesis. In this table, the absolute values are represented ($\log(X/H)_{\star}$+12) and the wavelengths are in \AA. }\\

\section{Results}
\label{sec:results}

\begin{figure}[t]
\centering
\includegraphics[scale=0.34]{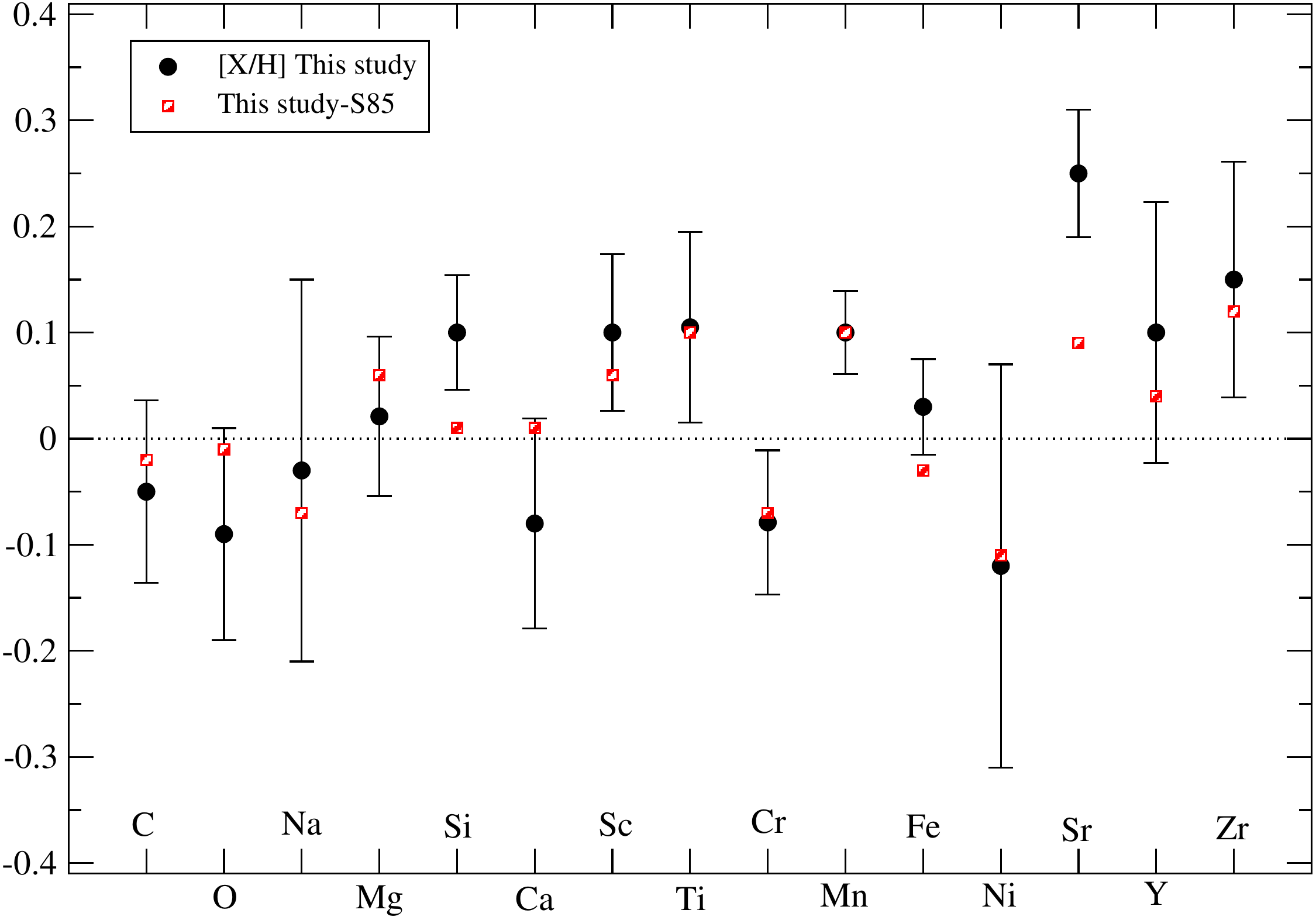}
\caption{Comparison of the abundances determined in this study and the one derived by Steffen 1985 (S85) for Procyon (red squares). The derived abundances are in black circles.}
\label{fig:procyon}
 \end{figure}

\textbf{We have first tested the spectrum synthesis on Procyon whose abundances are almost solar (Steffen 1985). For elements having lines on the 3 AURELIE spectral ranges, abundances agree weel. The derived abundances are displayed in Fig.~\ref{fig:procyon}. We have found nearly solar abundances for all the elements except for strontium. The differences between the abundances we derived and those derived by Steffen (1985) are depicted as squares in Fig.~\ref{fig:procyon}. They are less than 0.10 dex for 13 out of 15 elements and less than 0.15 dex for the remaining (Ni and Zr), typically less then the order of magnitude of the uncertainties. We found an apparent rotational velocity of 6 \kms\ and a microturbulent velocity of 2.2 \kms, in good agreement with Steffen's values ($v_{e}\sin i$=4.5 \kms \ and $\xi_{t}$=2.1 \kms, Steffen 1985)}. \\

\begin{figure*}[htb]
\vskip0.7cm
\caption{Abundance patterns for the "normal" A (a), Am (b) and F (c,d) stars of the Hyades cluster. A maximum $\pm$0.30 dex error bar is displayed. 
The horizontal dashed line represents the solar composition.}
\centering
\begin{tabular}{cc}
\includegraphics[scale=0.3]{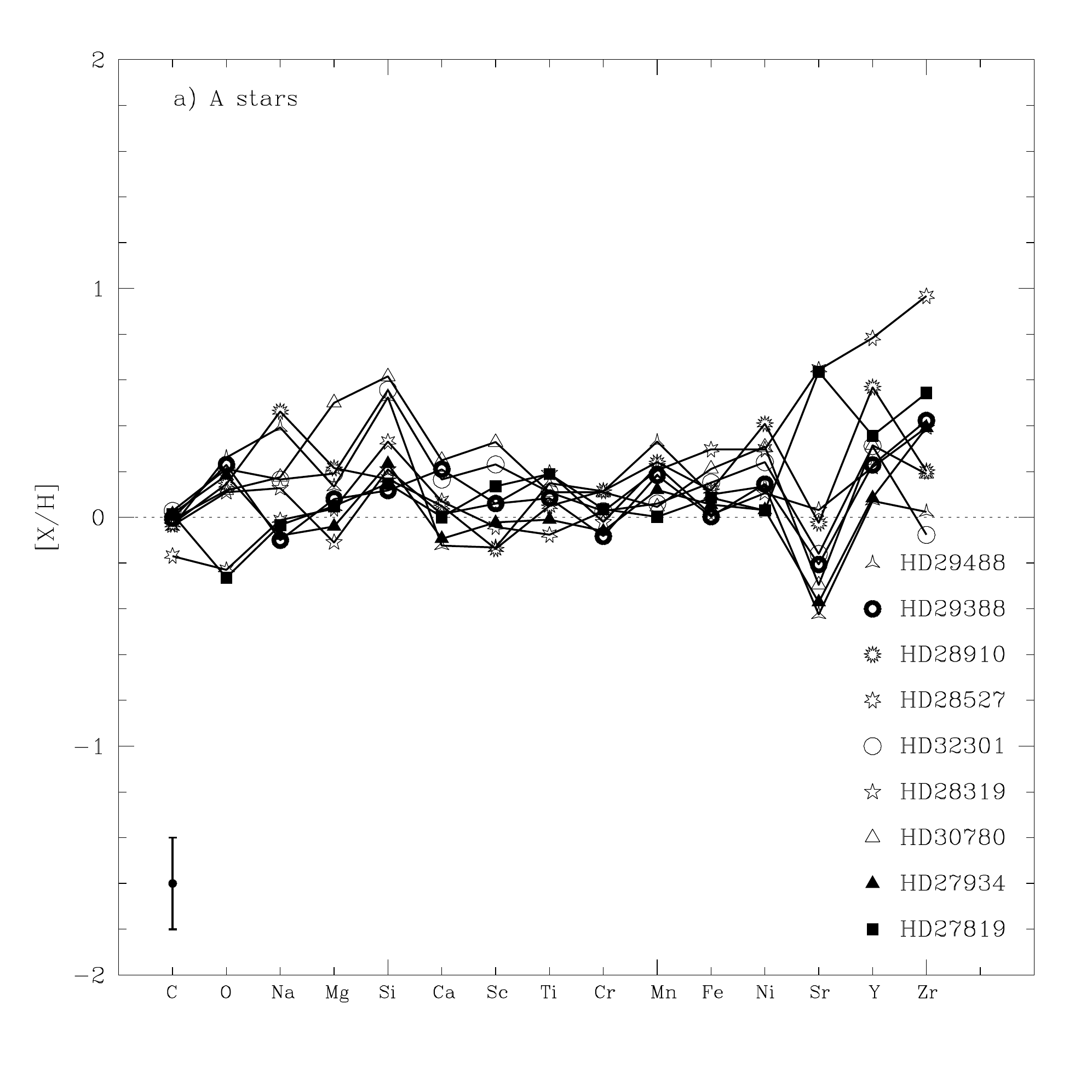}&
\includegraphics[scale=0.3]{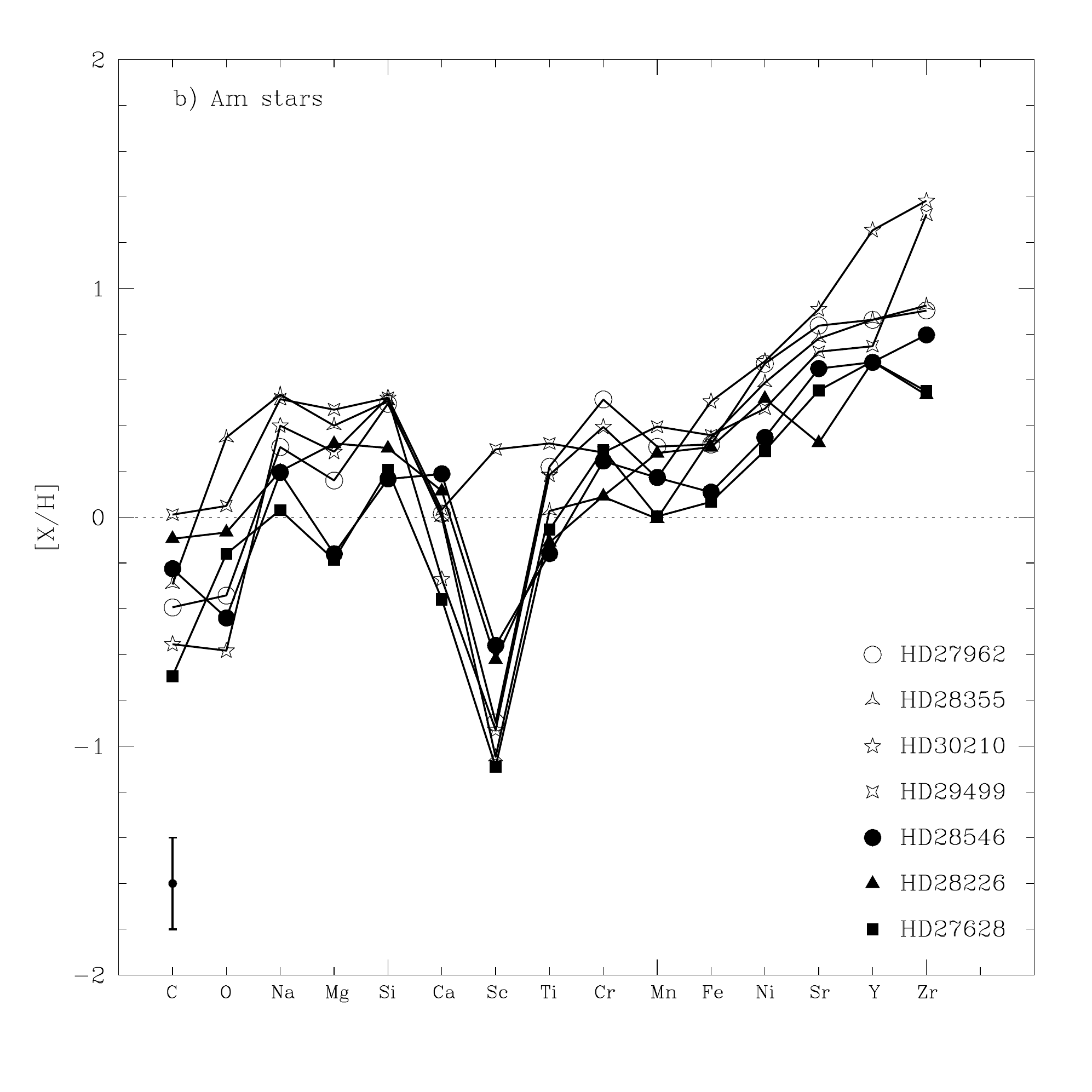}\\   
\includegraphics[scale=0.3]{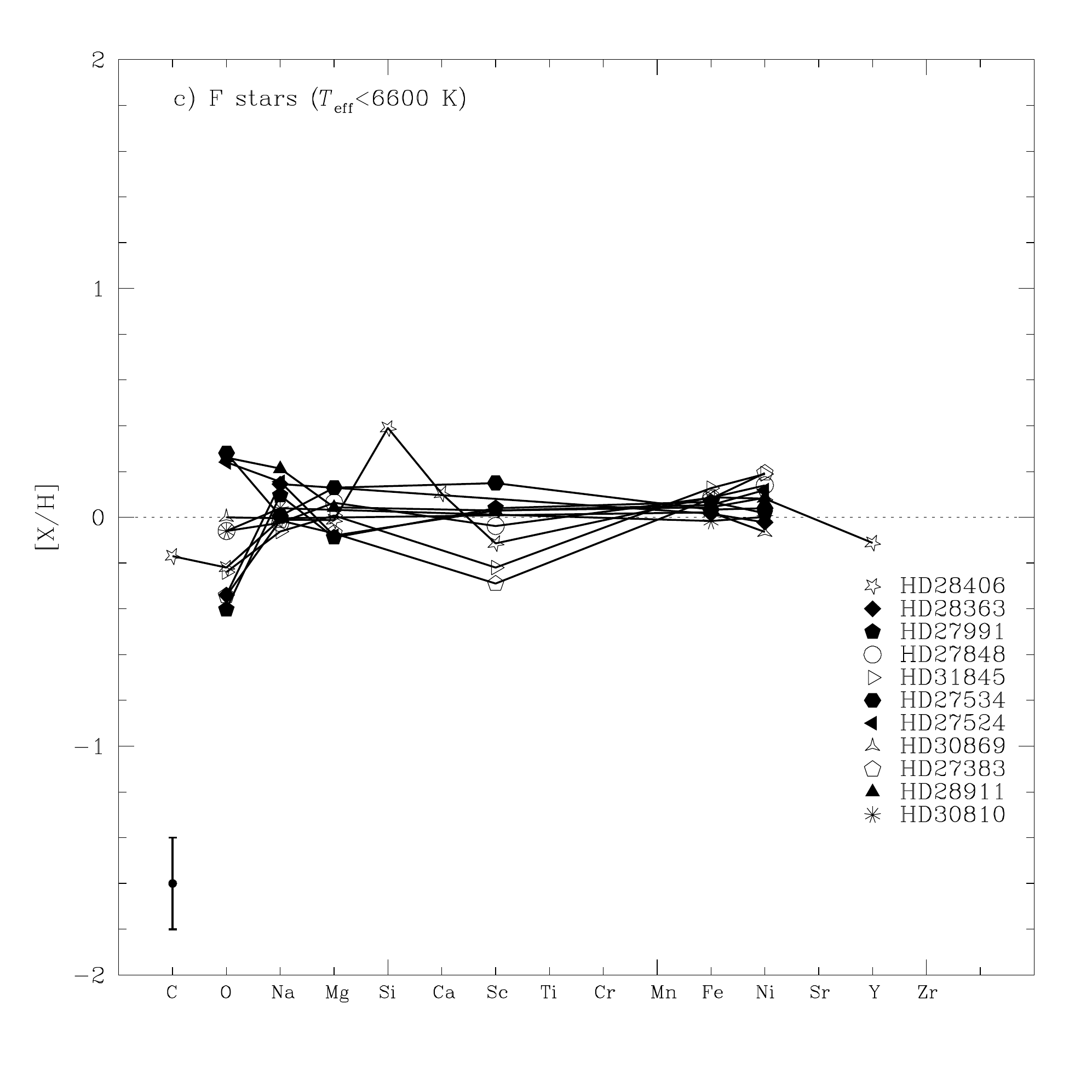}&
\includegraphics[scale=0.3]{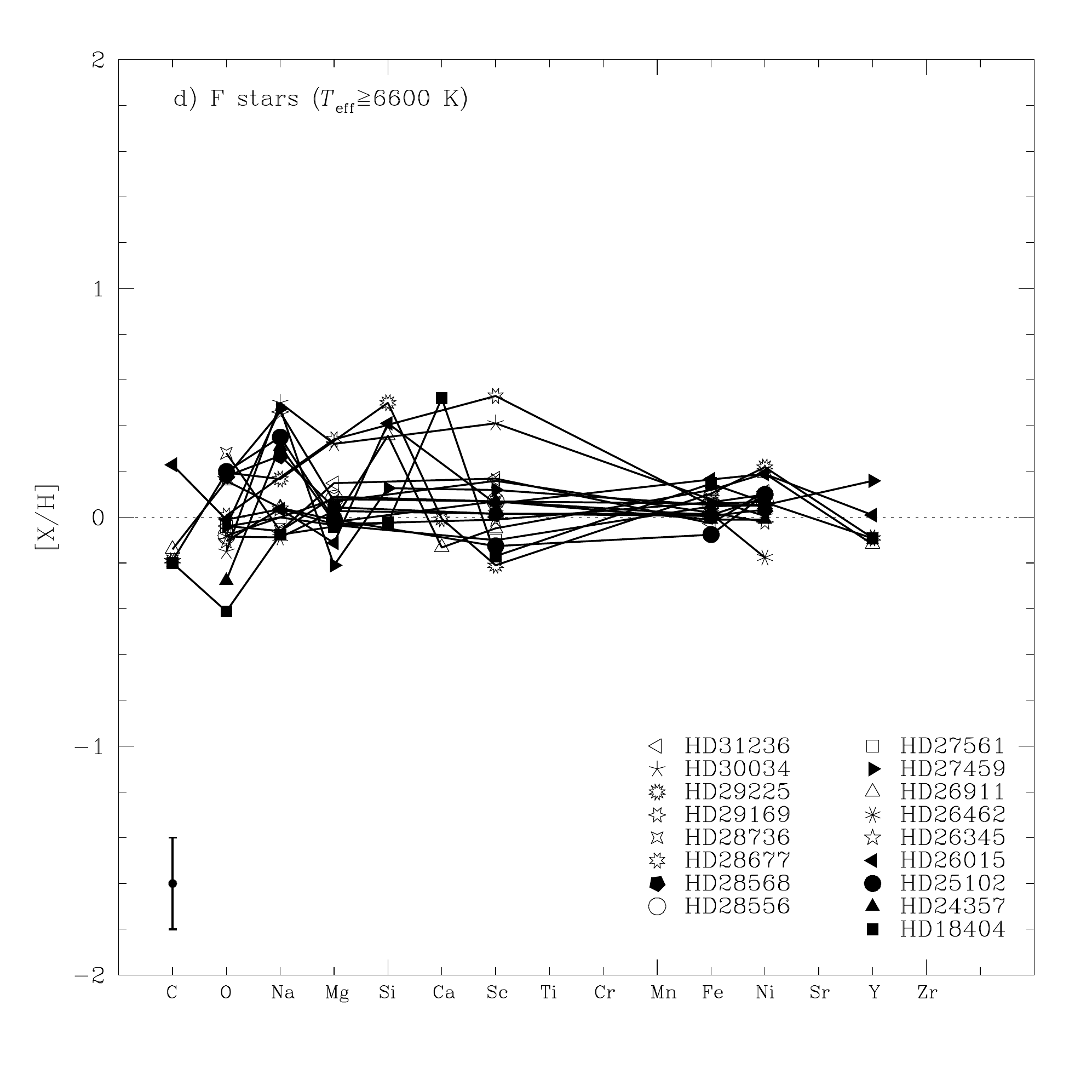}
\end{tabular}

\label{abond-A-Am-F}
 \end{figure*}

 \begin{figure*}
\vskip0.7cm
\centering
\includegraphics[width=0.7\linewidth]{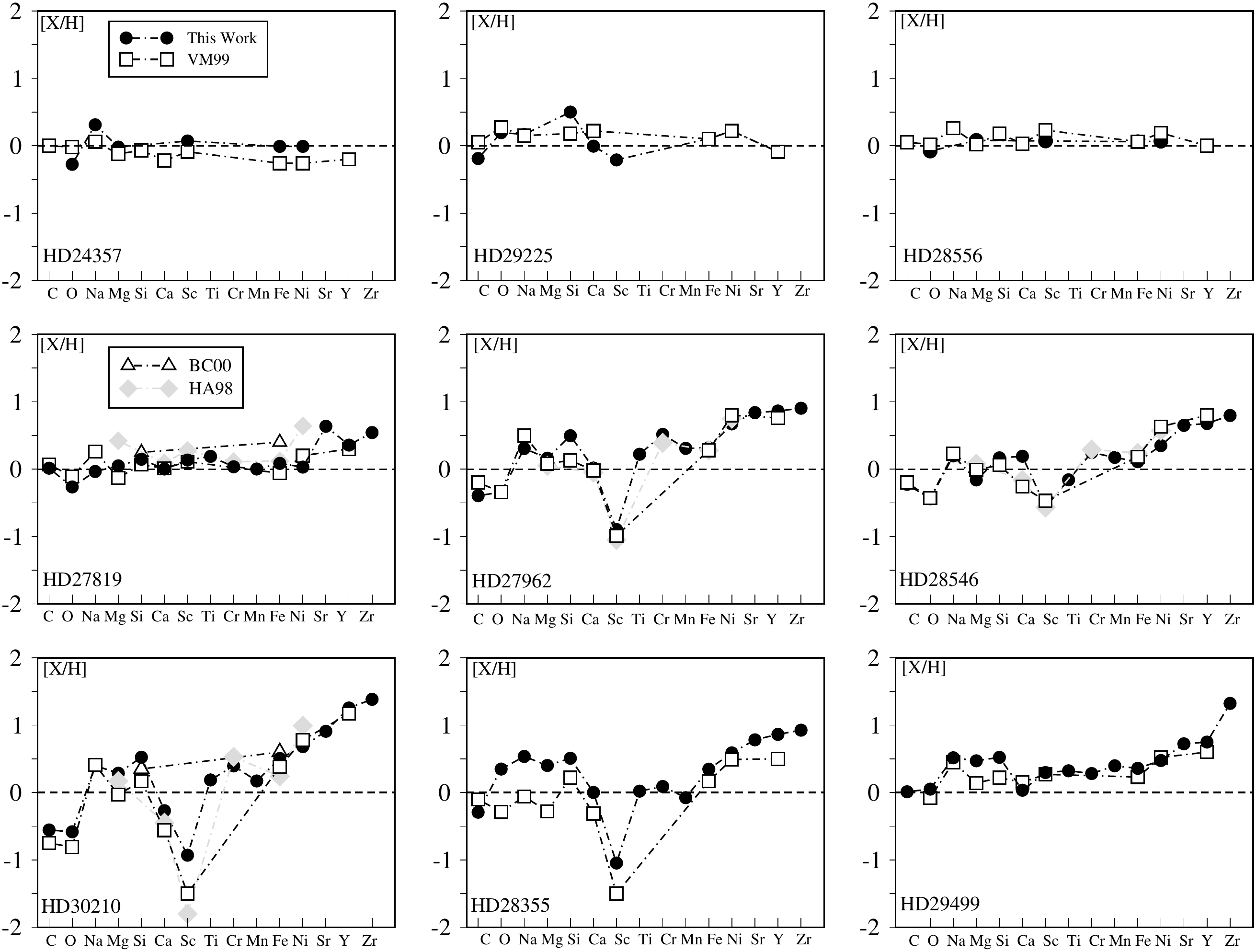}\\
\caption{Comparisons between the abundances derived in this work (filled circles), \vm \ (VM99, open squares),  Hui-Bon-Hoa \& Alecian (1998) (HA98, filled diamonds) 
and Burkhart \& Coupry (2000) (BC00, open triangles) for 9 stars of the Hyades cluster.}
\label{comp}
\end{figure*}

\subsection{Abundance patterns and comparison with previous studies}
\label{sec:trends}

Abundance patterns graphs where abundances are displayed against atomic number
$Z$ are particularly useful to compare the behaviour of A, Am and F stars for
different chemical elements. 
The abundance patterns for A, Am and F stars are displayed in Figs.
\ref{abond-A-Am-F}a-d. The pattern for A stars resembles that of 
the A stars in Coma Berenices and in the Pleiades (Papers I and II). 
Among the 9 A stars, the elements that exhibit the largest star-to-star variations
are Sr, Y and Zr (1.0 to 0.8 dex)
while C and Cr display the lowest variations (0.2 dex). The amplitudes of variations for
the other elements, O, Na, Mg, Si, Ca, Sc, Ti, Mn, Fe and Ni range from 0.25 to
0.60 dex (see further discussion of individual elements).\\
The 7 Am stars display the characteristic jig-saw pattern with larger excursions
around the solar composition than the A stars do. Almost all Am stars are
heavily deficient in Sc, but not all are deficient in Ca and the deficiencies
are more modest for this element. Almost all are more deficient in C and
O than the A stars, and all are more enriched in iron peak elements and heavy elements
(Sr and beyond). For all chemical elements, the
star-to-star variations of Am stars are usually larger than for the A
stars.\\
In contrast, F stars exhibit little scatter around the mean abundances. For clarity reasons, the abundances of  F stars are sorted out in two graphs: the data for stars cooler than 6600 K appear in Fig. \ref{abond-A-Am-F}c, and for those hotter than 6600 K in Fig. \ref{abond-A-Am-F}d. At the age of the Hyades, a F star with a 
temperature of 6600 K has a 1.3-1.4 M$_\odot$ mass. As explained in sect. \ref{evol-F}, the evolutionary models show that  the effects of 
atomic diffusion are more pronounced in all stars earlier than F5 (M$_\star >$1.3M$_\odot$) (Turcotte et al. 1998a). Sorting out the F stars into two groups (\teff$<$6600 K and \teff$\geq$6600 K) helps to highlight the occurrence of diffusion in the most massive F stars. \\
Graphically, we have compared the abundances derived in this study (filled
circles) with previous
determinations for 9 stars in Figure \ref{comp}. The abundances of Mg, Ca, Sc, Cr, Fe and Ni
derived by Hui-Bon-Hoa \& Alecian (1998) for HD27819, HD27962 and HD30210 are depicted
as losanges, the abundances of Si and Fe in HD27819 and HD30210 derived by \cite{burkhart-coupry-2000} 
as empty triangles and those of C, O, Na, Mg, Si, Ca, Sc, Fe, Ni and Y  derived by \vm\ as empty squares. 
For all stars, the overall shapes of the abundance patterns agree well.
Differences exist for individual elements mostly because of the use of different microturbulent velocities, rotational velocities, and in 
the case of A stars, different ionization levels. \\
We have also compared the derived iron abundances for F stars in this work with
the compilation available in Perryman et al. (1998) in
Table \ref{diff-Fstars}. The \xsurh{Fe} determinations come from \textbf{Chaffee et al. (1971) (CCS), 
Boesgaard \& Budge (1988) (BB), Boesgaard (1989) (B) and Boesgaard \& Friel (1990) (BF)}, they
mostly rely on adjustments of theoretical equivalent widths to observed ones. 
Differences arise from the usage of different effective temperatures with a fixed gravity ($\logg$=4.5 dex), older version of Kurucz ATLAS model atmospheres, 
different microturbulent velocities (determined using Nissen's 1981 fit) and different neutral iron lines.   

\subsection{Comments on particular stars}

Am stars are expected to be underabundant in light elements, underabundant in calcium and/or scandium as well as overabundant in iron-peak 
and heavy elements. HD27962 is the hottest A star in the Hyades, Mermilliod
(1982) suggested that it may be a blue straggler on basis of its location
on the HR diagram. Conti (1965) classified HD 27962 as an Am star based on the weakness of the scandium line at $\lambda$4246 \AA \  and the strength of strontium line 
at $\lambda$4215 \AA. Abt (1985) assigned a spectral type Am (A2KA3HA5M) to HD 27962.
Our analysis shows that scandium is deficient by -0.90 dex
and that iron-peak and heavy elements are enhanced in this star so that it has
the characteristics of an Am star.
Our analysis of HD28355 (A7V/A5m) also confirms its Am status as scandium is
deficient by -1.05 dex and 
iron-peak and heavy elements are enhanced. \cite{1977RMxAA...2..231H} had previously classified HD28355 as an Am star on
basis of Geneva photometry.  Both HD 27962 and HD 28355 are classified as Am in 
the catalog of Renson (1992). 
The seven stars represented in Fig. \ref{abond-A-Am-F}b are classified as
Am in the catalog of Renson (1992). The abundances of all these stars except HD
29499 display the characteristic jig-saw pattern of Am stars: underabundances of
light elements, of Ca and/or Sc and overabundances of metals and heavy elements.
Our abundance analysis 
strongly suggest that HD29499 (A5m) may actually be a normal A star: it does not
have Ca nor Sc deficiencies and is only moderately enriched in iron-peak and
heavy elements.Its apparent rotational velocity is around 60 \kms which is rather large for an Am star. 
\cite{1995ApJS...99..135A} have classified HD29499 as a giant star with 
metallic lines (A9III) but the surface gravity we found for HD 29499 suggests
that it still be on the Main Sequence. \\

\begin{table}[h]
 \caption{Iron abundances comparisons.}
 \large
 \centering
 \label{diff-Fstars}
\begin{tabular}{cccc}
\hline
Star& Reference& \xsurh{Fe}& \xsurh{Fe}$_{\rm{this \, work}}$ \\   \hline 
HD24357&BB&0.30&-0.014 \\ 
HD25102&BB&0.20&-0.075 \\
HD26015&BF&0.10&0.166 \\
HD26345&BF&0.18&0.065 \\
HD26462&BF&0.08&0.016 \\
HD26911&BB&0.27&0.120 \\
HD27383&CCS&0.23&0.082 \\
HD27561&BF&0.16&0.048 \\
HD27848&B&0.16&0.086 \\
HD27991&BF&0.11&0.067 \\
HD28406&BF&0.12&0.088 \\
HD28736&BB&0.13&0.009 \\
HD29225&BB&0.19&0.104 \\
HD30810&CCS&0.16&-0.016 \\
HD31845&BF&0.30&0.128 \\ \hline
\end{tabular}
   \end{table} 


\begin{table*}
 \caption{Mean abundances and dispersions.}
 \large
 \centering
 \label{abon-moyenne-deviation}
\begin{tabular}{c|ccc|ccc|ccc}
\hline
Elements&F stars&$\sigma_{F}$ & (Max-Min)$_{[\frac{X}{H}]}$ &A stars &  $\sigma_{A}$& (Max-Min)$_{[\frac{X}{H}]}$& Am stars & $\sigma_{Am}$& (Max-Min)$_{[\frac{X}{H}]}$\\   \hline 
$[\frac{C}{H}] $	&-0.09  &0.16&0.43   &-0.15  &0.21&0.72 &-0.32&0.27  &0.71 \\  
$[\frac{O}{H}] $	&-0.03	&0.21&0.69   &-0.03  &0.27&0.93 &-0.17&0.30  &0.93	    \\  
$[\frac{Na}{H}]$	&0.11	&0.18&0.59   &0.20   &0.20&0.64 &0.31&0.23   &0.50	  \\  
$[\frac{Mg}{H}]$	&0.03	&0.11&0.55   &0.15   &0.21&0.68 &0.18&0.21   &0.66	 \\  
$[\frac{Si}{H}]$	&0.29   &0.18&0.52   &0.35   &0.17&0.50 &0.39&0.17   &0.36    \\  
$[\frac{Ca}{H}]$	&0.12   &0.24&0.65   &0.02   &0.16&0.61 &-0.04&0.17  &0.55      \\ 
$[\frac{Sc}{H}]$	&0.03	&0.18&0.82   &-0.27  &0.49&1.42 &-0.69&0.64  &1.39	  \\  
$[\frac{Ti}{H}]$	&-      &-    &-     &0.08   &0.13&0.48 &0.06 &0.13  &0.48      \\  
$[\frac{Cr}{H}]$	&-      &-   &-      &0.14   &0.16&0.60 &0.27&0.21   &0.42      \\  
$[\frac{Mn}{H}]$	&-      &-   &-      &0.17   &0.12&0.47 &0.18&0.13   &0.47      \\  
$[\frac{Fe}{H}]$	&0.05	&0.05&0.24   &0.19   &0.14&0.50 &0.29&0.17   &0.44    \\   
$[\frac{Ni}{H}]$	&0.07	&0.09&0.40   &0.33   &0.21&0.65 &0.51&0.26   &0.39	 \\   
$[\frac{Sr}{H}]$	&-      &-    &-     &0.29   &0.46&1.33 &0.68&0.61   &0.58      \\  
$[\frac{Y}{H}]$		&-0.04  &0.10 &0.28  &0.54   &0.32&1.18 &0.82&0.42   &0.57      \\  
$[\frac{Zr}{H}]$	&-       &-    &-    &0.60   &0.41&1.46 &0.93&0.53   &0.83      \\ \hline

\end{tabular}
   \end{table*}
   
\subsection{Behavior of the abundances of individual elements}
\label{discussion}
\begin{figure}[t]
\centering
\includegraphics[width=\linewidth]{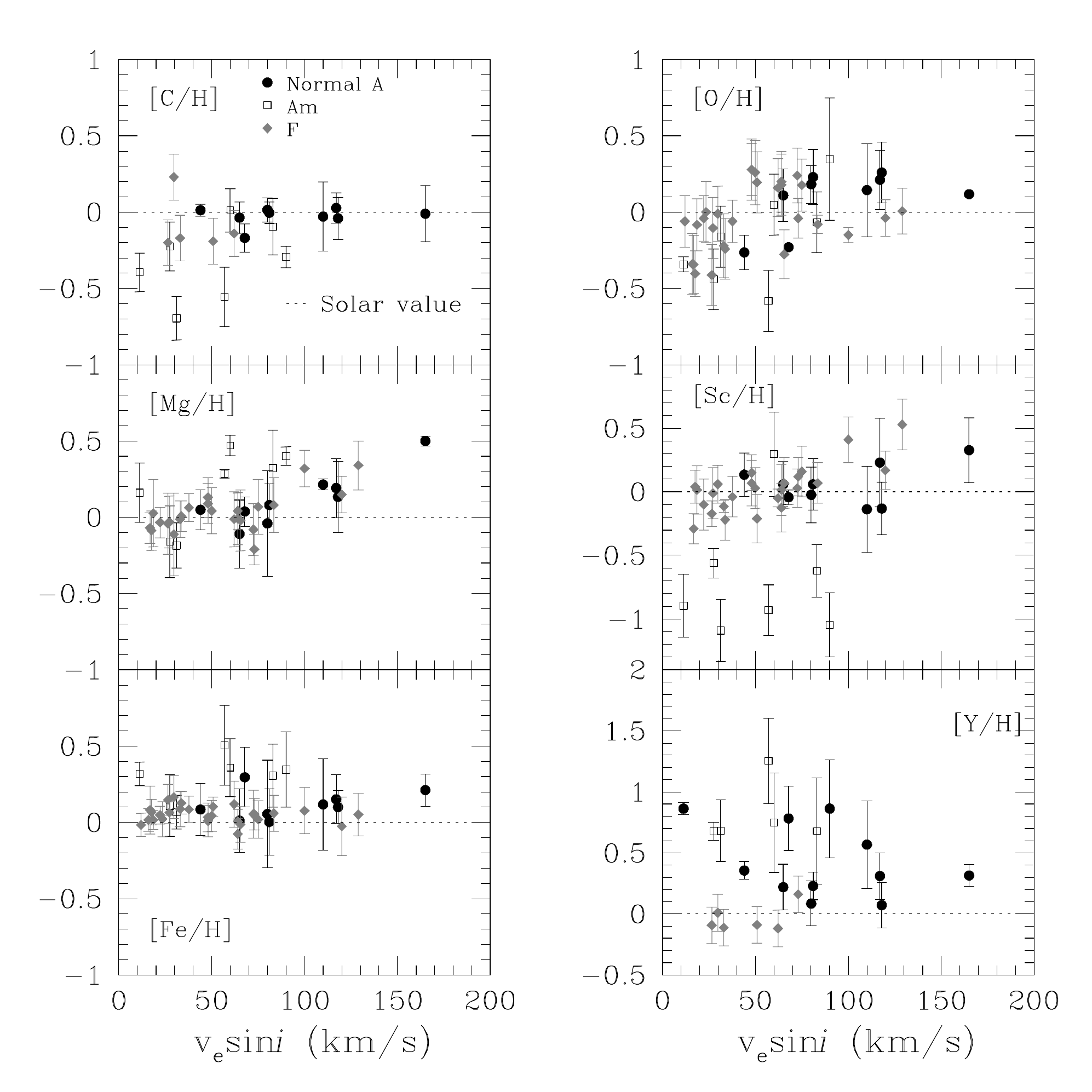}\\
\caption{Abundances of C, O, Mg, Sc, Fe and \textbf{Y} versus $v_{e}\sin i$ for A,Am and F stars.}
\label{abund-vsini}
\end{figure}

\begin{figure}[]
\centering
\vskip0.5cm
\includegraphics[width=\linewidth]{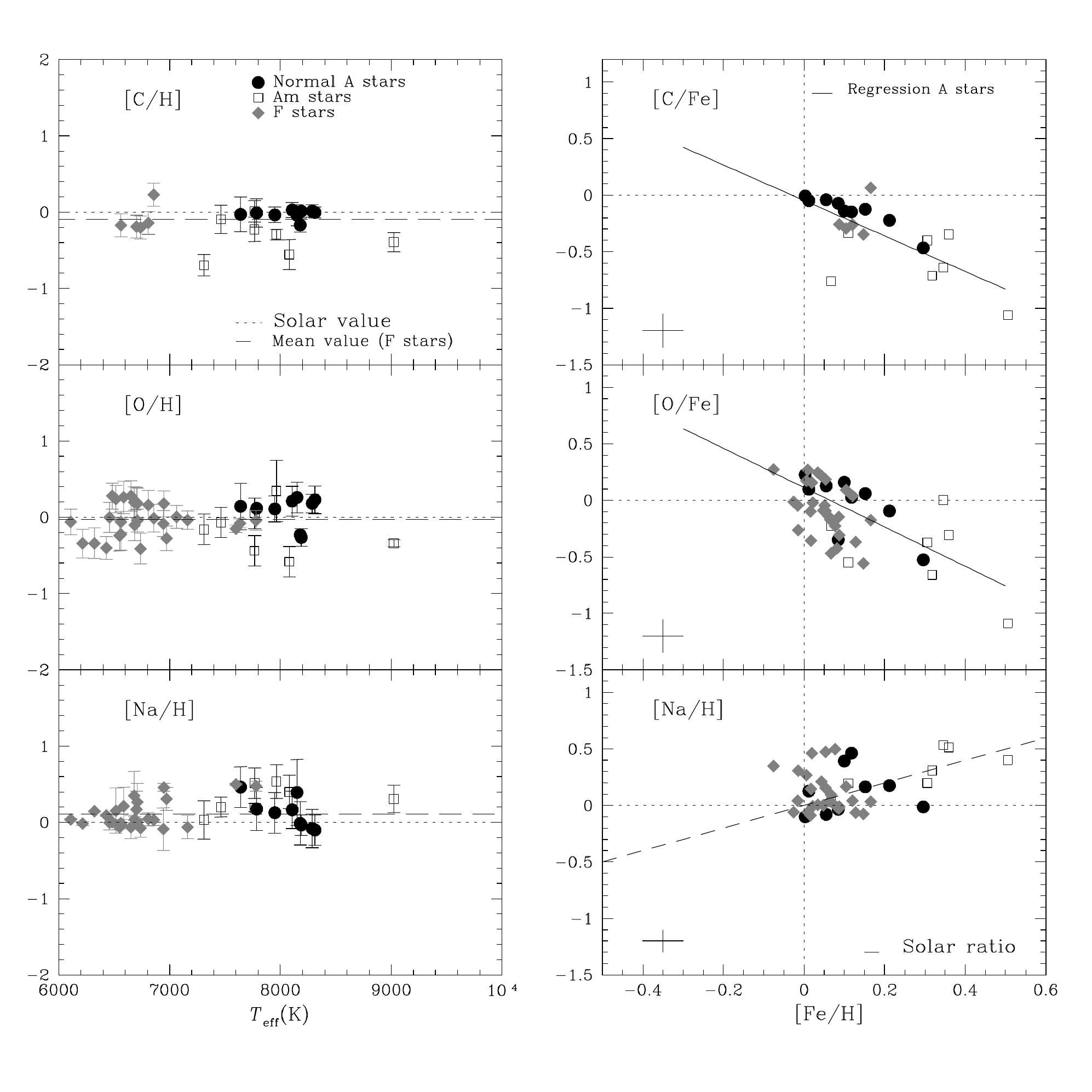}\\
\caption{Left panel: Abundances of carbon, oxygen and sodium versus effective temperature. The dotted line corresponds to the solar value and the 
dashed line to the mean value determined for the F stars of the cluster. Right panel: [C/Fe], [O/Fe] and [Na/H] versus [Fe/H]. The filled dots 
correspond to normal A stars, the open squares correspond to Am stars and the filled diamonds correspond to F stars. In the plot representing [Na/H] versus [Fe/H], 
the dashed line corresponds to the solar [Na/Fe] ratio. The error bars in the right panel represent the mean standard deviation for the displayed abundances.}
\label{CCA}
\end{figure}

\begin{figure}
\centering
\vskip0.5cm
\includegraphics[width=\linewidth]{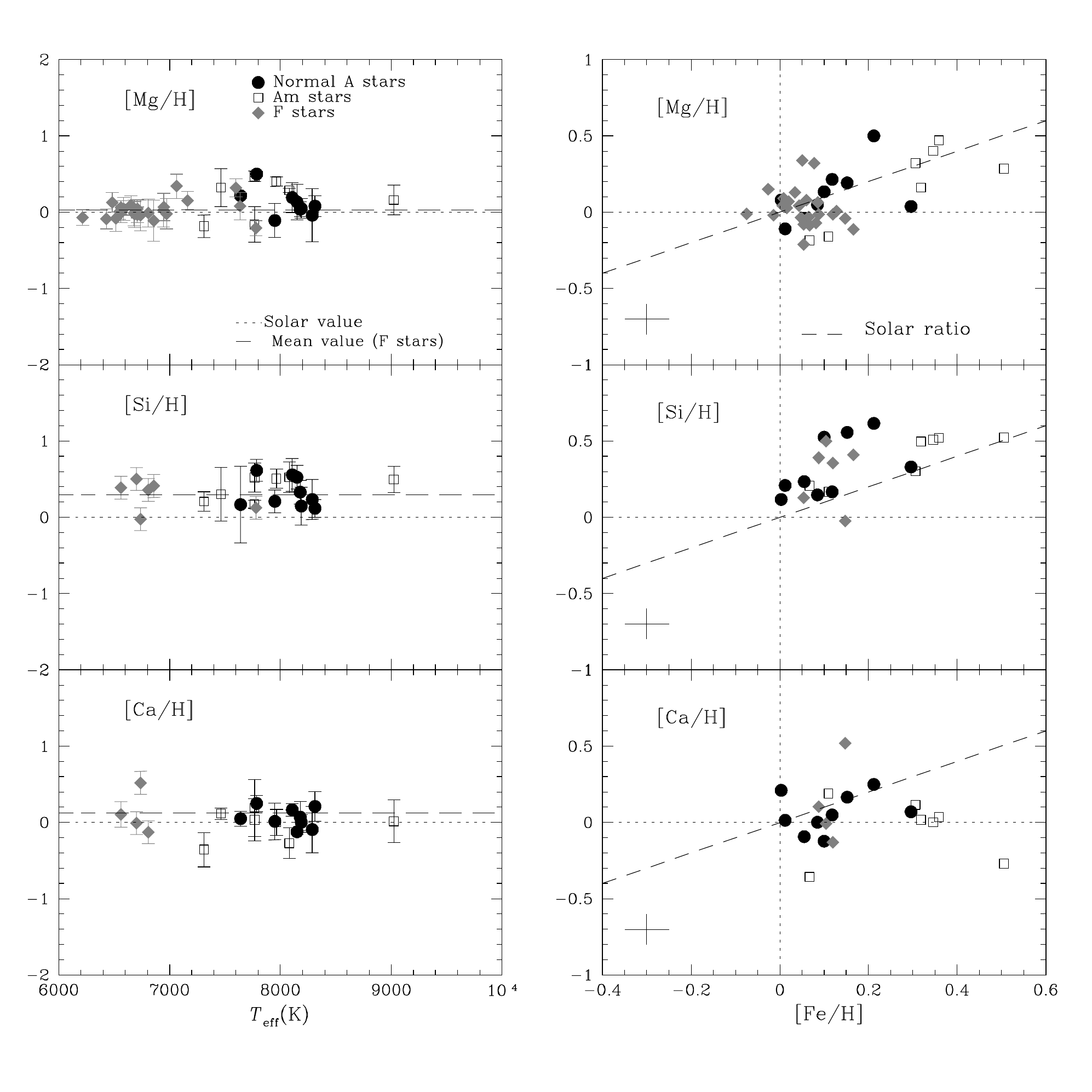}
\caption{Left panel: Abundances of magnesium, silicon and calcium versus effective temperature. The dotted line represents 
the solar value and the dashed one represents the mean abundance of F stars. Right panel: [Mg/H], [Si/Fe] and [Ca/H] versus [Fe/H]. 
The symbols are the same as in Figure \ref{CCA}. The dashed lines represent the solar ratios.}
\label{MgSiCa}

\end{figure}

\begin{figure}[t!]
\centering
\vskip0.5cm
\includegraphics[width=\linewidth]{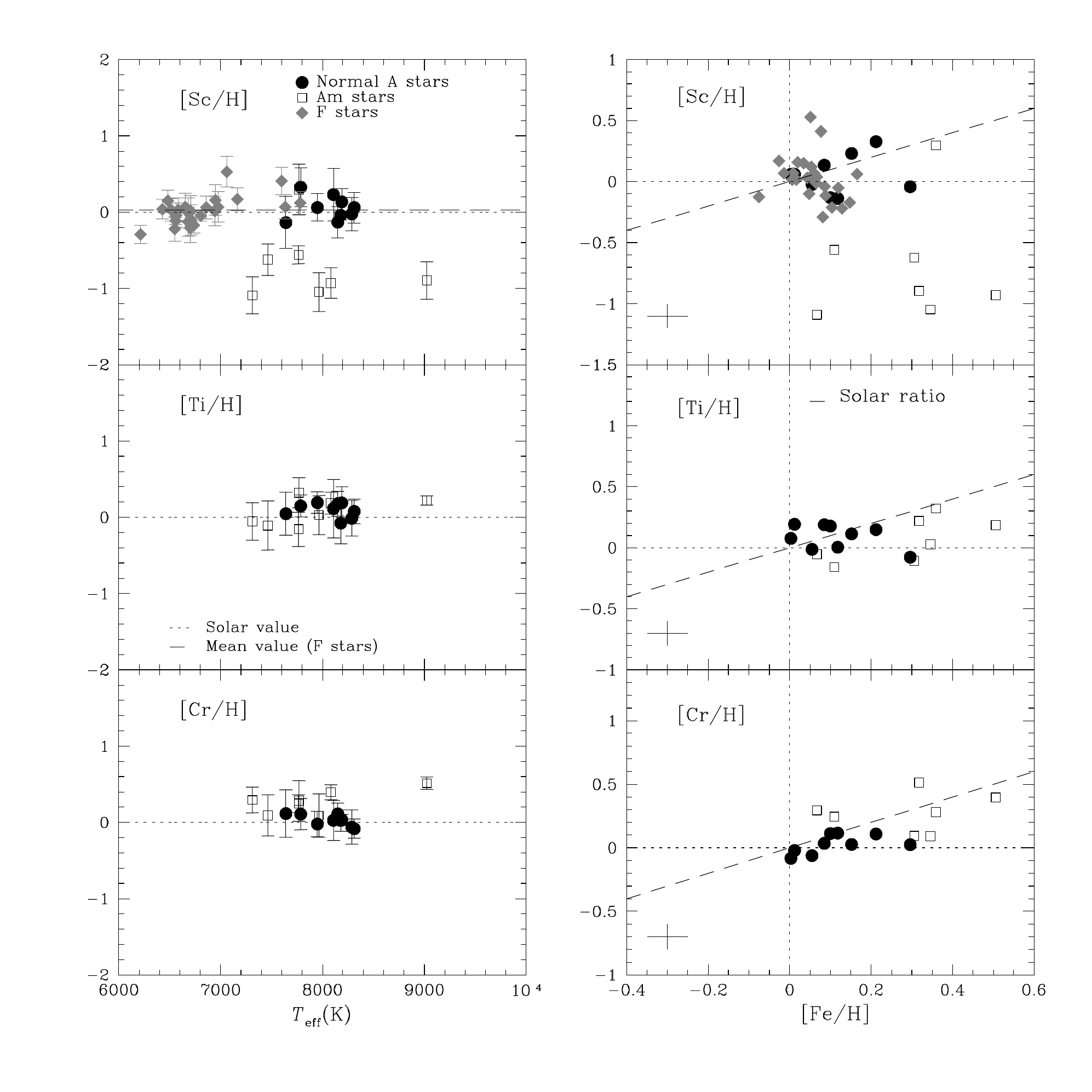}
\caption{Left panel: Abundances of scandium, titanium and chromium versus effective temperature. The dotted line represents the solar 
value and the dashed one represents the mean abundance of F stars. Right panel: [Sc/H], [Ti/Fe] and [Cr/H] versus [Fe/H]. The symbols are the 
same as in Figure \ref{CCA}. The dashed lines represent the solar ratios.}
\label{ScTiCr}
\end{figure}

\begin{figure}
\centering
\vskip0.5cm
\includegraphics[width=\linewidth]{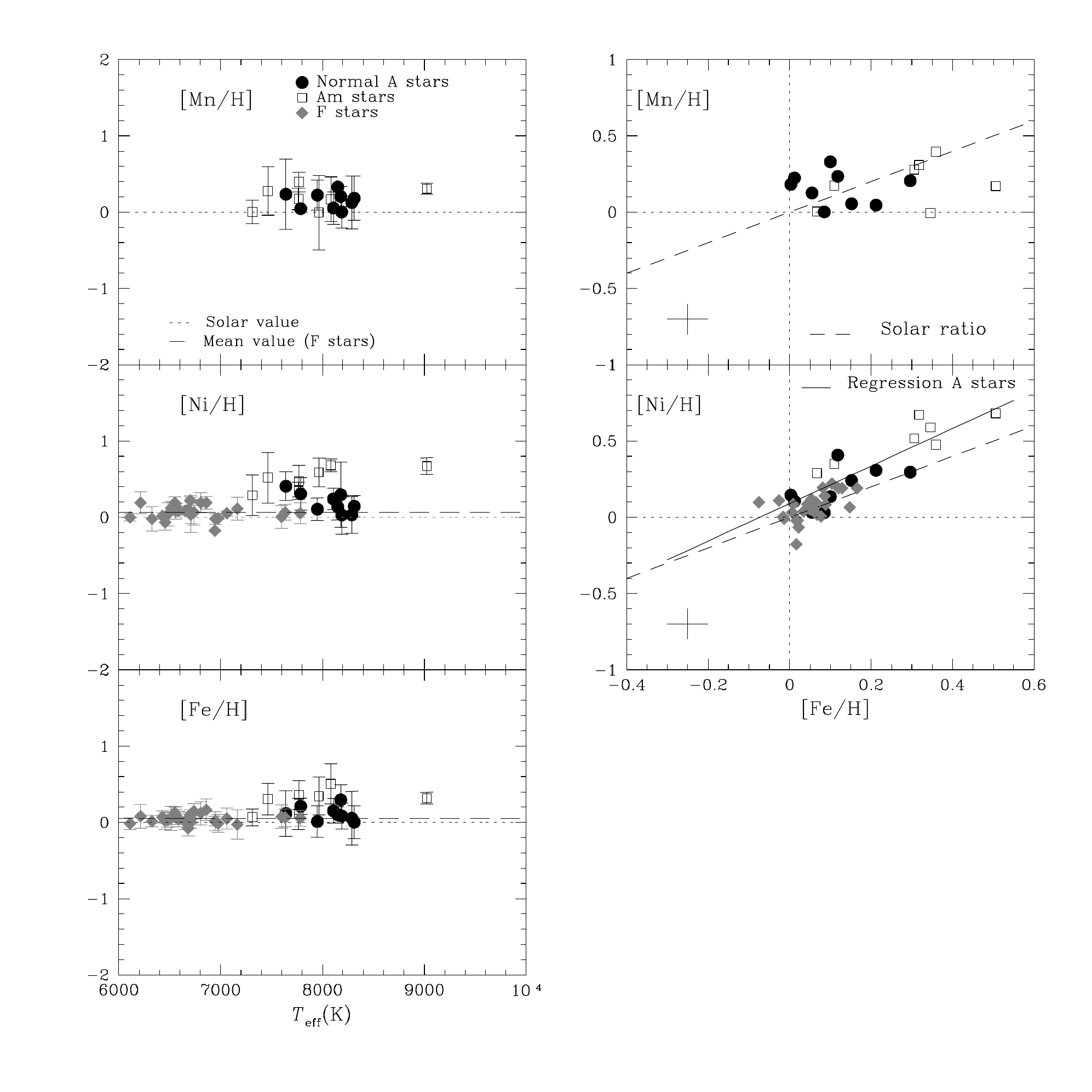}
\caption{Left panel: Abundances of manganese, iron and nickel versus effective temperature. The dotted line represents the solar 
value and the dashed one represents the mean abundance of F stars. Right panel: [Mn/H] and [Ni/H] versus [Fe/H]. The symbols are the 
same as in Figure \ref{CCA}. The dashed lines represent the solar ratios.}
\label{MnFeNi}
\end{figure}

\begin{figure}
\centering
\vskip0.5cm
\includegraphics[width=\linewidth]{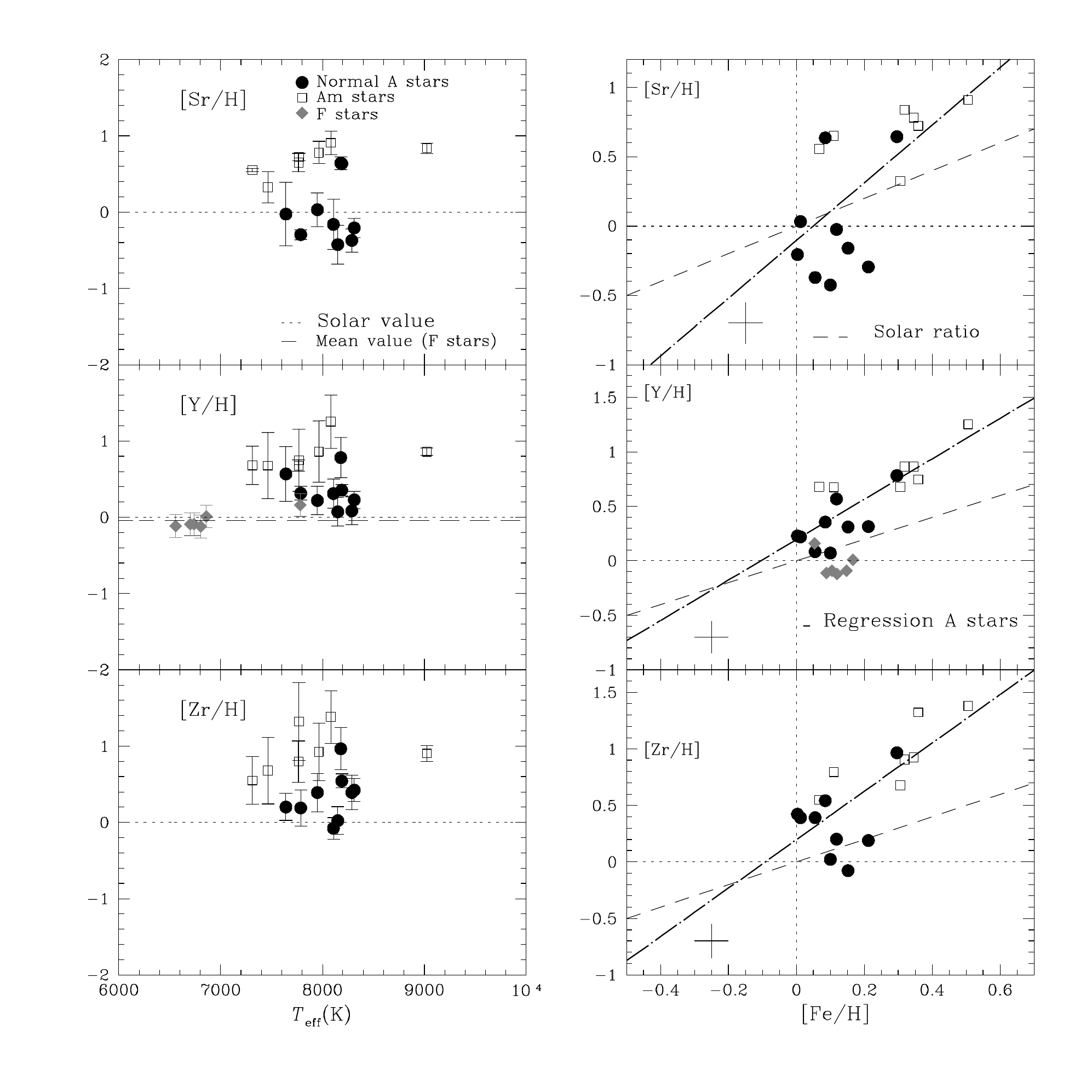}
\caption{Left panel: Abundances of strontium, yttrium and zirconium versus effective temperature. Right panel: 
[Sr/H], [Y/Fe] and [Zr/H] versus [Fe/H]. The symbols are the same as in Figure \ref{CCA}.}
\label{SrYZrBa}
\end{figure}

As we have done for the Coma Berenices cluster (see Paper I), the behavior of
the found abundances has been studied versus 
apparent rotational velocity (\vsini) and effective temperature (\teff). 
Any correlation/anticorrelation would be very valuable to theorists
investigating the various hydrodynamical mechanisms affecting photospheric
abundances.
As we emphasized in Paper I, the existence of star-to-star variations with
fundamental parameters can be established independently of errors in the
absolute values of the oscillator strengths, since all stars will be affected in
the same manner.\\ 
Second, we have searched whether the abundances of individual elements correlate
with that of iron. \textbf{We expect the abundances of Fe, Ti, O, Cr, Mg, Mn, C, Ca and Ni to be fairly reliable as we synthesized several lines of quality A to D for these elements. For Y and Zr, several lines are available but their accuracy is unknown, the abundances of these elements should therefore be taken with caution. The abundances of Sr, derived from 2 transitions whose oscillator strengths have unknown inaccuracies, are likely to be inaccurate.}
\\
Abundances are displayed against \teff\ in the left part of Figs. \ref{CCA} to
\ref{SrYZrBa}. Inspection of these figures reveal that there is no systematic
slope (positive nor negative) and that for a large number of chemical 
elements, A stars display star-to-star variations in abundances larger than the typical uncertainty. 
Table \ref{abon-moyenne-deviation} presents the mean abundance, the standard deviation and maximum spread for all chemical elements in \textbf{F, A (normal and Am) and Am stars}. Scatter around the mean value is more important in A stars than in F stars, namely for C, Na, Sc, Fe, Ni, Sr, Y and Zr. This behavior was
already found in Coma Berenices and the Pleiades (see Papers I and II). \\
The abundances of C, O, Mg, Sc, Fe and \textbf{Y} are displayed versus \vsini \ for A, Am
and F stars in Fig. \ref{abund-vsini}. For a given element, there is usually a considerable scatter in abundances at a given rotation rate. None of the derived abundances in this study exhibits a clear correlation nor anticorrelation with \vsini.
\cite{1991ApJ...370..693C} have analyzed the effect 
of meridional circulation on the chemical separation of elements in rotating
stars. They showed that for stars rotating at less than \vsini$\leq$100 \kms, no correlation 
should be expected between abundances and apparent rotational velocities. This
prediction is verified by our findings (see Fig. \ref{abund-vsini}). 
Even if we distinguish two velocity regimes (\vsini$\leq$100 \kms \ and
vsini$\ge$100 \kms), we fail to find any dependence between the abundances of any of the \textbf{15} chemical elements and the 
apparent rotational velocity.  Recently, Takeda et al. (2008) have found that the 
peculiarities (underabundances of C, O, and Ca) seen in
slow rotators efficiently decrease with an increase of rotation and almost 
disappear at \vsini$\geq$100 \kms. We confirm that for these chemical elements
abundance anomalies vanish at \vsini$\geq$100 \kms .\\

\textbf{Carbon and oxygen abundances display large star-to-star variation in A stars. No clear correlation was found between the abundances of C or O and \xsurh{Fe}\bf . However, both \xsury{C}{Fe}\bf \ and \xsury{O}{Fe}\bf \ are anticorrelated with \xsurh{Fe}\bf .  For the carbon lines used in our study, non-LTE abundance corrections for the A stars (7000 K$<$\teff$<$10000 K) are expected to be negative (Rentzsch-Holm 1996) and do not affect the star-to-star dispersion in \xsurh{C}\bf . Non-LTE corrections for oxygen abundances are negligible in the case of the OI lines considered here (see Paper I). Carbon and Oxygen tend to be more deficient in Am stars than in A stars of similar effective temperatures or rotation rate.\\
For A stars, sodium abundances appear to be slightly correlated to the iron abundances as seen in Fig. \ref{CCA}. We found important star-to-star abundance variations in  \xsurh{Na}\bf .  In Fig. \ref{MgSiCa}, the scatter of \xsurh{Mg}\bf \ for both F and A stars does not exceed the typical uncertainty (0.20 dex) which suggests that there are no significant star-to-star variations in magnesium abundances. As mentioned in Paper I, the MgII $\lambda$4481 \AA \ triplet yields higher
abundances than other MgII lines, therefore we excluded this line from our analysis. The corrected \xsurh{Mg}\bf \ abundances appear to be slightly correlated with \xsurh{Fe}\bf . The ratio \xsury{Mg}{Fe}\bf \ is close to solar for the A stars. The silicon lines synthesized in this work have low quality oscillator strengths and have not been updated for the recent values of the $\log gf$ of the NIST\footnote{http://www.nist.gov/} database. The systematic found overabundances could be due to incorrect oscillator strengths, therefore the Si abundances should be viewed with caution. There does not seem to be significant star-to-star variations in \xsurh{Si}\bf .\\
Calcium abundance does not exhibit real star-to-star variations neither any clear correlation with repsect to \xsurh{Fe}\bf. Not all Am stars are underabundant in calcium, only two of the 7 Am stars exhibit large underabundances in Ca. This result differs from Varenne \& Monier's (1999) findings because of the use of different microturbulent velocities and ionisation level as explained in Sec.\ref{sec:trends}. In \vm, the derived microturbulent abundances were larger (up to 5 \kms) that those derived here, leading to lower abundances of calcium. Star-to-star variations in scandium abundance are clearly present for A stars (Fig. \ref{ScTiCr}). Scandium is the most scattered of all analyzed elements. Scandium abundances do not appear to be correlated to iron abundances for both A and F stars. All Am stars except for HD 29499 are deficient in scandium and fall in the lower right part of Figure \ref{ScTiCr}. The two stars, HD27962 and HD28355, for which we confirm the status as Am stars, are located in that region. \\
Titanium, chromium and manganese abundances were derived for A stars only since none of the Ti, Cr and Mn lines observed with AURELIE have oscillator strengths accurate enough for abundance determinations. There does not seem to be significant star-to-star variation  in \xsurh{Ti}\bf, \xsurh{Cr}\bf \ nor \xsurh{Mn}\bf \ for the A stars (Fig.\ref{ScTiCr}). The titanium abundances of the Hyades A stars do not appear to be correlated with the iron abundances. The chromium abundance appears to be only loosely correlated to that of iron.\\
Iron abundances have been derived for all F and A stars of our sample. Neutral iron lines were used for F stars yielding a mean abundance of $< \xsurh{Fe} >_{\rm{F}}$=0.05$\pm$0.05 dex. This value, which represents the average metallicity of the cluster, is almost 0.1 dex smaller than the 
value derived by Boesgaard \& Friel (1990) from a different sample of 14 F Dwarfs (+0.127$\pm$0.022 dex using different FeI lines from ours). For the 16 A stars, 27 lines of FeII were synthesized. 
The normal A and Am stars scatter around their mean abundance with a maximum spread of about
0.50 dex, which is more than twice larger than the typical uncertainty on \xsurh{Fe}\bf \ is about 0.20 dex. This suggests real star-to-star variations in \xsurh{Fe}\bf \ among the Hyades A stars. Nickel behaves similarly to iron (see Fig. \ref{MnFeNi}), the A stars display large star-to-star variations in \xsurh{Ni}\bf \ and the abundances are clearly correlated with the iron abundances, the correlation coefficient being close to 1.\\
All Am stars appear to be overabundant in strontium. Star-to-star variation in  \xsurh{Sr}\bf \ are clearly present. Strontium  abundances are only loosely correlated with that of iron (right part of Fig. \ref{SrYZrBa}). Yttrium and zirconium are found overabundant in all A stars with a real star-to-star variations in \xsurh{Y}\bf \ and \xsurh{Zr}\bf \ (Figure \ref{SrYZrBa}). As for strontium, yttrium and zirconium abundances are correlated to \xsurh{Fe}\bf \ but appear to increase more rapidly than \xsurh{Fe}\bf\ (slope of 1.8 for Y and 2.1 for Zr).\\}

\section{Discussion}
\label{sec:models}

\subsection{Self-consistent evolutionary models}

The derived abundances have been compared to the predictions of recent evolutionary models. These models are calculated with the Montr\'eal stellar evolution code. 
The chemical transport problem is treated with all known physical processes from first principles, which includes radiative accelerations, 
thermal diffusion and gravitational settling (for more details see Turcotte et al. 1998b, Richard et al. 2001 and references therein). 
These models follow the chemical evolution of most elements as well as some isotopes up to $Z\leq28$ (28 species in all). As the abundances change, 
the Rosseland opacity and radiative accelerations are continuously recalculated at each mesh point and for every time step during evolution which means that the 
treatment of particle transport is completely self-consistent. The spectra used to calculate the monochromatic opacities are taken from the OPAL database (Iglesias et al. 1996). 
The radiative accelerations are calculated as described in Richer et al. (1998) with corrections for the redistribution of momentum from Gonzalez et al. (1995) and LeBlanc et al. 
(2000). The mixing length parameter and initial helium abundance ($\alpha=2.096 $ and $ Y_0=0.2779$ respectively) are calibrated to fit the current luminosity and radius of the 
Sun (see Turcotte et al. 1998b, model H). Models are evolved from the pre-main sequence with a solar scaled abundance mix. 
The initial abundance ratios are given in Table 1 of Turcotte et al. (1998b). For the initial mass fraction of metals we used both $Z_0=0.02$, 
the solar metal content, and $Z_0=0.024$ (to represent the increased metallicity of the Hyades, Lebreton et al. 2001).

\begin{figure}[ht!]
\centering
\includegraphics[scale=0.4]{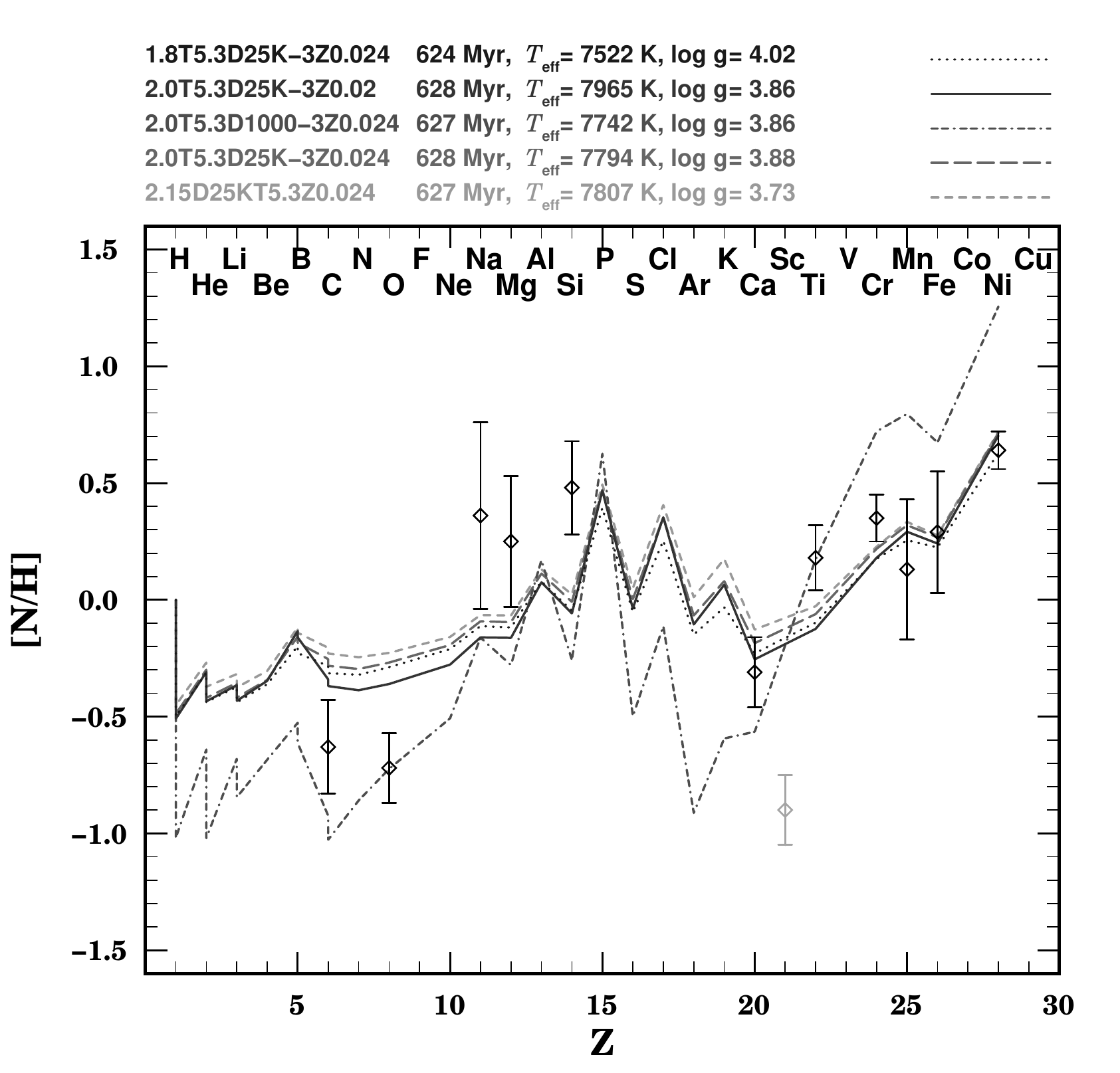}
\caption{Comparison of the predicted surface abundances for stars of 1.8, 2.0 and 2.15\,M$_\odot$ with different turbulence prescriptions and different initial metallicities with those derived (depicted as diamonds with their respective uncertainties) for the Am star HD30210 ($T_{\rm eff}=8082$\,K and $\log g=3.92$). The gray symbol for Sc indicates that it is not considered in the evolutionary models.}
\label{Am30210}
\end{figure}

\subsubsection{The A stars}
\label{evol-A}
The derived abundances in A stars are compared to the predictions of self-consistent evolutionary models calculated as in Richer et al. (2000). 
These models include an arbitrary parameter, the amount of mass mixed by
turbulent mixing, in order to lower the effects of chemical separation and
better fit the abundances 
of AmFm stars. The effect of turbulence is to extend the mixed mass below the surface convection zone, down to layers where atomic diffusion is less efficient 
(ie. where time scales are longer due to higher local density). {\it A priori}, the cause of turbulence is not considered, however it was found to be compatible 
with rotationally induced turbulence as calculated by Zahn 2005 and Talon et al. 2006. The surface abundances are shown to depend solely on the amount of mass mixed 
as well as initial metallicity.
In Figure \ref{Am30210}, we compare the predicted surface abundances for five
models with masses of 1.8, 2.0 and 2.15\,M$_\odot$ to the observed abundances of the Hyades A2m star 
HD30210 ($T_{\rm eff}=8082$\,K and $\log g=3.92$\,dex). The models are shown for both Z=0.02 and Z=0.024. 
To help interpret the chosen nomenclature, the 2.0T5.3D1000-3 corresponds to a 2.0\,M$_\odot$ star for which the abundances are completely homogenized down to $\log T=5.3$. 
Below this point, as we move deeper inside the star, the turbulent diffusion coefficient $D_{\rm T}= 1000\,D_{\rm He}$ (the helium diffusion coefficient) decreases 
as $\rho^{-3}$ (2.0T5.3D25K-3 would have the same behavior except that the turbulent coefficient $D_{\rm T}= 25000\,D_{\rm He}$ at $\log T=5.3$). 
It is clear from the plot that the amount of prescribed turbulence has an influence on the surface behavior. 
The model 2.0T5.3D1000-3 does not sufficiently impeed microscopic diffusion to reproduce most abundances. However, for all other models with greater turbulence, 
the predicted carbon and oxygen underabundances are not as important as the observed anomalies. The predicted slight underabundances of Na, Mg and Si are 
not observed in HD30210, but are observed in a few of the Am stars. However the silicon abundance is probably too large. 
The model that best fits the data is 2.15T5.3D25K-3Z0.024, which has a turbulence prescription which was equivalently able to reproduce the observations in 
Coma Berenices (see Figures 14 and 15 of Paper I). As in Paper I, models with less turbulence roughly replicate the observations for $Z\,<\,12$ and the 
more turbulent models are better able to reproduce the heavier elements ($Z\,>\,15$). The different metallicities do not lead to significant differences 
in the predicted abundance patterns, but do have an important effect on $T_{\rm eff}$. 

\subsubsection{The F-type stars}
\label{evol-F}
For the F dwarfs, we have compared the found abundances to the predicted surface
abundances for C, O, Na, Mg, Fe and Ni using Turcotte's evolutionary models at
620 Myr (Turcotte et al. 1998a). These models, calculated for masses ranging
from 1.1-1.5 $\rm{M}_{\odot}$, treat radiative diffusion in detail but do not
include macroscopic mixing processes (meridional circulation, turbulence or mass loss). 
They show that the effects of atomic diffusion, namely the appearance of surface abundance anomalies, can be expected in all stars earlier than 
F5 (M$_*\,>\,$1.3 $\rm{M}_{\odot}$). \\
The found mean carbon abundance for F stars, $<$\xsurh{C}$>$=-0.09 dex with very
small dispersion, does not agree with predicted underabundance at 620 Myr ($\log\rm{age}=8.79$ 
in Figure 7 of \cite{turcotte-et-al-98}) and for a 1.4$\rm{M}_{\odot}$ star,
representative of the F stars analysed here.
The oxygen abundances, which show large scatter for the F stars, can typically
be underabundant by -0.41 dex or overabundant by up to 0.28 dex for a 
1.4$\rm{M}_{\odot}$ F star (roughly an effective temperature of 6700 K at the
age of the Hyades). Again, they do not agree with the predicted surface
underabundances \xsurh{O}\ predicted by Turcotte et al. (1998a).
The predicted solar Na abundances match reasonably well of our determinations
for F stars. 
Magnesium is predicted to be slightly underabundant (Fig. 7 of Turcotte et al.
1998a), whereas for most F stars we find overabundances. 
Finally, iron and nickel are found to be mildly overabundant in case of 
F stars with \teff$\in$[6600K,6800K] ($<$\xsurh{Fe}$>$=0.06 dex  and $<$\xsurh{Ni}$>$=0.11 dex).
These results disagree with the overabundances of 0.5 and 0.8 dex respectively for a 1.4\,M$_\odot$. As expected, these purely diffusive models typically predict too little C and O and too much iron-peak elements.

\section{Conclusion}
\label{sec:conclusion}
Selected high quality lines in new high resolution \'echelle spectra of 16 A and 28 F stars of the Hyades have been synthesized in a uniform manner to derive LTE abundances, a few of which have been corrected for Non-LTE effects whenever possible. Even when binary stars are removed, the abundances of several chemical elements for A stars and early F stars exhibit real star-to-star variations, significantly larger than for the late F stars. The largest spreads occur for Sc, Sr, Y, Zr while the lowest are for Mg, Si and Cr for A stars. Gebran et al (2008) and Gebran \& Monier (2008) had already found similar behaviour in the Coma Berenices and the Pleiades.
The derived abundances do not depend on effective temperatures nor apparent rotational velocities as expected since the timescales of diffusion are much shorter than those of rotational mixing (Charbonneau \& Michaud 1991). The abundances of Cr, Ni, Sr, Y and Zr are correlated with the iron abundance as was found for the Pleiades and Coma Berenices. The ratios [C/Fe] and [O/Fe] are anticorrelated with [Fe/H] (particularily true for normal A stars). Compared to normal A stars, all Am stars in the Hyades appear to be more deficient in C and O and more overabundant in elements heavier than Fe but not all are deficient in calcium and/or scandium. The F stars have nearly solar abundances for almost all elements except for Si and Ca.
The Blue Straggler HD 27962 appears to have abundances characteristic of an Am star (scandium deficiency and enrichment in iron-peak and heavy elements). Conversely our abundance analysis of HD 29499 (A5m) yields normal abundances in Ca and Sc and only moderate enrichment in iron-peak and heavy elements suggesting that this star might be a normal A star.
The detailed modelling of the A2m star HD 30210 including radiative diffusion and different amounts of turbulent diffusion reproduces the overall shape of the abundance pattern for this star but not individual abundances. Models with the least turbulence reproduce the abundances of the lightest (Z $<$ 12) and those with most turbulence reproduce abundances of elements with Z $>$ 15. For a few elements, the discrepancies between derived and predicted abundances could be due to Non-LTE effects. However, the inclusion of competing processes such as different prescriptions of rotational mixing (Zahn 2005) and/or different amounts of mass loss (Vick et al. , in preparation) could well improve the agreement between observed and predicted abundance patterns.

\begin{acknowledgements}
We warmly thank the OHP night staff for the support during the observing runs. This research has used the SIMBAD, WEBDA, VALD, NIST and Kurucz databases. 
MV thanks the D\'epartement de physique at l'Universit\'e de Montr\'eal as well as the GRAAL at l'Universit\'e Montpellier II for financial support and the R\'eseau Qu\'eb\'ecois de Calcul de Haute Performance (RQCHP) for providing us with the computational resources required for this work. 
A special thanks to Georges Michaud and Olivier Richard for their careful reading of the manuscript and useful suggestions. LF received support from the Austrian Science Foundation (FWF project P17890-N2).

\end{acknowledgements}

\Online

\begin{table*}
\caption{Abundances relative to hydrogen and to the solar value, for A stars.}
\label{abondances-A}
\centering 
\begin{tabular}{cccccccccccc} 
\hline
HD   	   & SpT 	    &CI	  &$\sigma_{C}$	   &OI   &$\sigma_{O}$	 & NaI   &$\sigma_{Na}$   &MgII   &$\sigma_{Mg}$    &SiII &$\sigma_{Si}$ \\ \hline \\
HD27628  & A3m   & -0.69  & 0.14  &  -0.16   &  0.04   &    0.03	 &  0.25   &    -0.18  &   0.15  &   0.21	 &  0.13	 \\
HD27819  & A7V   & 0.01   & 0.04  &  -0.26   &  0.11   &    -0.03   	&  0.13   &    0.05   &   0.13  &   0.15	 &  0.25	 \\
HD27934  & A7V   & 0.01   & 0.08  &  0.18    &  0.13   &    -0.08   	&  0.25   &    -0.04  &   0.34  &   0.23	 &  0.26	\\
HD27962  & A2IV  & -0.39  & 0.13  &  -0.34   &  0.05   &    0.31	 &  0.18   &    0.16   &   0.19  &   0.50	 &  0.17	 \\
HD28226  & A3m	 & -0.09  & 0.19  &  -0.07   &  0.06   &    0.20	 &  0.13   &    0.32   &   0.25  &   0.30	 &  0.35	\\
HD28319  & A7III & -0.04  & 0.10  &  0.11    &  0.18   &    0.13	 &  0.27   &    -0.11  &   0.22  &   0.21	 &  0.15	\\
\end{tabular}
 \end{table*}

\begin{table*}[htb!]
\caption{Abundances relative to hydrogen and to the solar value, for F stars.}
\label{abondances-F}
\centering 
 \begin{tabular}{cccccccccccccc} 
\hline
HD   	   & SpT 	    &CI	  &$\sigma_{C}$	   &OI   &$\sigma_{O}$	 & NaI   &$\sigma_{Na}$   &MgI   &$\sigma_{Mg}$    &SiII    
&$\sigma_{Si}$ &CaII        &	$\sigma_{Ca}$\\ \hline \\
HD18404  &   F5IV   &   -0.20   &  0.16   &   -0.41   &    0.20   &   -0.08   &  0.12   &  -0.04   &  0.20  &  -0.02  &  0.14   &  0.52  &      0.17	\\
HD24357  &   F4V    &		 &	    &	-0.28   &  0.16   &    0.31	&  0.15   &  -0.02   &  0.20  &	    &	       &	 &	      \\
HD25102  &   F5V    &		 &	    &	0.200	 & 0.20   &    0.35	&  0.32   &  -0.01   &  0.17  &	    &	       &	 &	      \\
HD26015  &   F3V    &	0.23	 &  0.15   &	-0.01   &  0.18   &    0.03	&  0.07   &  -0.11   &  0.22  &  0.41   &  0.18   &	 &	      \\
HD26345  &   F6V    &		 &	    &	-0.10   &  0.20   &    0.04	&  0.04   &  -0.03   &  0.17  &	    &	       &	 &	      \\
HD26462  &   F4V    &		 &	    &	-0.08   &  0.17   &    -0.09	&  0.28   &  0.03    &  0.21  &	    &	       &	 &	      \\
\hline \hline
\end{tabular}
 \end{table*}

\begin{table*}
\caption{Absolute parameters for the observed stars.}
\label{tab:LTMA}
\centering
\begin{tabular}{ccccccccc}
\hline  \hline

Star&	L/L$_{\odot}$	&$\sigma_{\rm{L}}$&	$\log T_{\rm{eff}}$ &$ \sigma_{\log T_{\rm{eff}}}$& Mass (M$_{\odot}$)& $\sigma_{\rm{M}}$&Fr. age& $\sigma_{\rm{Fr. age}}$ \\ \hline
HD27628& 0.97&    0.05 &   3.864 &  0.007 &  1.71 &   0.12 &   0.36  &  0.05  \\
HD27819& 1.31&    0.05 &   3.913 &  0.007 &  2.05 &   0.10 &   0.63  &  0.09  \\
HD27934& 1.54&    0.05 &   3.918 &  0.007 &  2.28 &   0.16 &   0.83  &  0.13  \\
HD27962& 1.47&    0.05 &   3.955 &  0.006 &  2.27 &   0.23 &   0.83  &  0.13  \\
HD28226& 0.97&    0.05 &   3.873 &  0.007 &  1.73 &   0.09 &   0.37  &  0.06  \\
HD28319& 1.89&    0.05 &   3.900 &  0.007 &  2.58 &   0.18 &   1.05  &  0.16  \\
HD28355& 1.23&    0.05 &   3.901 &  0.007 &  1.96 &   0.10 &   0.55  &  0.08  \\
\hline									    
 \hline									    
\end{tabular}								    
\end{table*}

\newcommand{\logs}[1]{\ensuremath{\mathrm{log}(\mathrm{#1}/\mathrm{H})_{\star}}}

\newpage
\longtab{8}{

 \centering
 
\begin{longtable}{cccccccc}
\caption{Elemental abundances for each line of each chemical element for A and F stars. }
\label{element-abundances}

\endfirsthead
\caption{continued.}
\endhead

\endfoot

\hline 
\multicolumn{8}{c}{\Large{\bf{A stars}}}\\
\multicolumn{8}{c}{\bf{HD27628}}\\ \\
$\lambda$C&\logs{C}&$\lambda$O&\logs{O}&$\lambda$Na&\logs{Na}&$\lambda$Mg&\logs{Mg}\\ \hline
4932.049&	7.774&5330.700 &	 8.700&4668.559 &   5.900&4390.510&	  7.207 \\
5052.167&	7.759&6158. 1&	 8.560&4982.813 &   6.547&4481.1  &	  7.503 \\
5380.337&	7.605&&&6154.226 &   6.386&&			 \\
5793.120&	8.005&&&		       6160.747 &   6.456&&			 \\
\hline
 \end{longtable}}  
    
\end{document}